\documentclass[aps,pra,floatfix,reprint, superscriptaddress]{revtex4-1}

\usepackage{graphicx}
\usepackage{color}
\usepackage{xcolor}
\usepackage[normalem]{ulem}
\usepackage{cancel}
\usepackage{hyperref}
\usepackage{amsmath}

\begin{document}

\title{Overcoming the efficiency-bandwidth tradeoff for optical harmonics generation using nonlinear time-variant resonators}

\author{Maxim~R.~Shcherbakov}
\affiliation{School of Applied and Engineering Physics, Cornell University, Ithaca, NY 14853, USA}
\affiliation{Faculty of Physics, Lomonosov Moscow State University, Moscow 119991, Russia}

\author{Pavel~Shafirin}
\affiliation{Faculty of Physics, Lomonosov Moscow State University, Moscow 119991, Russia}

\author{Gennady~Shvets}
\affiliation{School of Applied and Engineering Physics, Cornell University, Ithaca, NY 14853, USA}

\date{\today}

\begin{abstract}
Highly resonant photonic structures, such as cavities and metasurfaces, can dramatically enhance the efficiency of nonlinear processes  by utilizing strong optical field enhancement at the resonance. The latter, however, comes at the expense of the bandwidth. Here, we overcome such tradeoff by utilizing time-varying resonant structures. Using harmonics generation as an example, we show that the amplitude and phase format of the excitation, as well as the time evolution of the resonator, can be optimized to yield the strongest nonlinear response. We find the conditions for an efficient synthesis of electromagnetic signals that surpass the cavity bandwidth, and discuss a potential experimental realization of this concept.
\end{abstract}

\maketitle

Resonant photonic structures are among the most common and fundamental building blocks in optics, finding numerous applications in sensing \cite{Luchansky2012}, optical computing \cite{Reiserer2015}, and metrology \cite{Aasi2013}. Particularly desirable are resonators with high quality ($Q$) factors because they provide large optical field enhancement, thereby enhancing the efficiency of nonlinear processes such as harmonics generation (HG)~\cite{Shcherbakov2014,Yang2015a,Vampa2017,Sivis2017}. However, high-$Q$ resonances limit the operational spectral range of such structures. For example, nonlinear generation of ultra-short high-harmonic pulses is prevented by the long filling time of the resonator ($\tau_{\rm res}\sim Q/\omega$, where $\omega$ is the resonance frequency)~\cite{Tsakmakidis2017,Mann2019}, during which a near-monochromatic harmonic signal is generated inside the resonator.

This limitation cannot be remedied by employing broadband excitation pulses  because only a small fraction of the laser spectrum effectively interacts with the resonator. Fortunately, this time-bandwidth limit holds only for as long as the cavity parameters---eigenfrequency and lifetime---are constant in time \cite{Xu2007d}. Here, we show that time-variant resonators (TVRs) whose resonant frequencies $\omega \equiv \omega(t)$ are varied in time can overcome these limitations, thereby resolving one of the most fundamental tradeoffs in nonlinear optics: the one between efficiency and spectral bandwidth.

TVRs have recently attracted considerable attention because of their potential for efficient frequency conversion \cite{Preble2007,Tanabe2009,Lee2018,Shcherbakov2019}, non-reciprocal devices \cite{Shaltout2015,Sounas2017,Caloz2018a}, topological photonics \cite{Minkov2016} and time-variant metamaterials \cite{Shaltout2015,Rogov2018,Qu2018a}. Using a simple theoretical framework based on the coupled mode theory (CMT) \cite{Haus1984,Fan2003a}, a variety of exotic phenomena have been successfully predicted, such as perfect frequency conversion into a sideband with light reversal \cite{Minkov2017}, continuous adiabatic frequency conversion \cite{Preble2007}, and many others. To our knowledge, the possibility of substantially modifying nonlinear light-matter interactions using TVRs has not yet been considered. In this contribution, we demonstrate that the nonlinear interaction between high-$Q$ TVRs and broadband laser pulses can be dramatically enhanced by employing frequency-chirped pulses. Specifically, we show that, for a given fluence and bandwidth, the interaction of judiciously engineered chirped laser pulses with nonlinear TVRs can produce much stronger and spectrally broader harmonics than their transform-limited counterparts interacting with either static or time-varying nonlinear resonators.

The rest of the paper is organized as follows. First, we introduce a CMT model of a single-mode TVR and derive an analytic expression for the mode's amplitude and phase in response to an arbitrary input pulse. Next, under the assumption of a linearly changing resonance frequency of the resonant mode, we derive the optimal driving pulse's parameters (its chirp and duration) designed to optimize harmonically generated (HG) signal inside the resonator. The fractional bandwidth of the HG signal is shown to be proportional to that of the laser pulse, $\Delta\omega_L/\omega_L\gg Q^{-1}$, where $Q$ is the quality factor of the mode. Finally, we discuss the implications of our findings to resonant generation of high optical harmonics and outline potential experimental realizations of broadband nonlinear interactions with high-Q TVRs.

\begin{figure}[b]\label{fig1}
\includegraphics[width=\columnwidth]{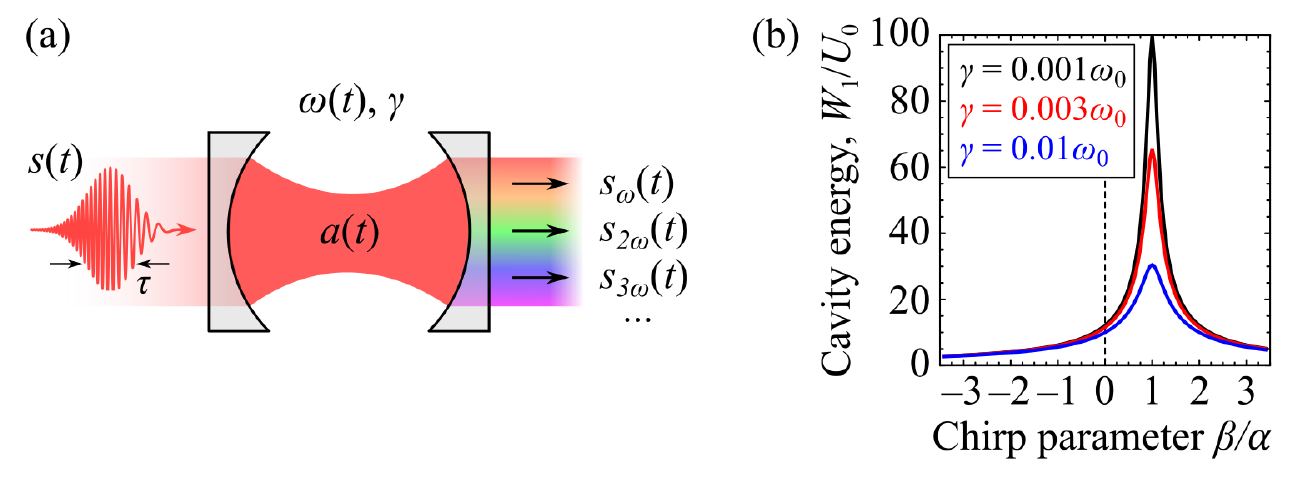}
\caption{(a) Schematic of a single-mode time-varying resonator (TVR) with a modal frequency $\omega(t)= \omega_0 + \alpha \omega_0^2 t$ and the decay rate $\gamma$. Input: frequency-chirped Gaussian pulsed source $s(t)$, see Eq.~(\ref{eq:OMC}). Output: fundamental and $n$'th harmonics: $s_{n\omega}\propto a^n(t)$. (b) Time-integrated normalized energy $W_1/U_0$ inside the TVR as a function of the linear frequency chirp $\beta$. Pulse duration: $\tau=100/\omega_0$, normalized rate of change of the modal frequency: $\alpha=10^{-3}$, phase of the excitation pulse: $\psi(t) =\omega_0 t(1 + \beta \omega_0 t/2)$.
\label{fig1}}
\end{figure}

\begin{figure*}\label{fig2}
\includegraphics[width=1.2\columnwidth]{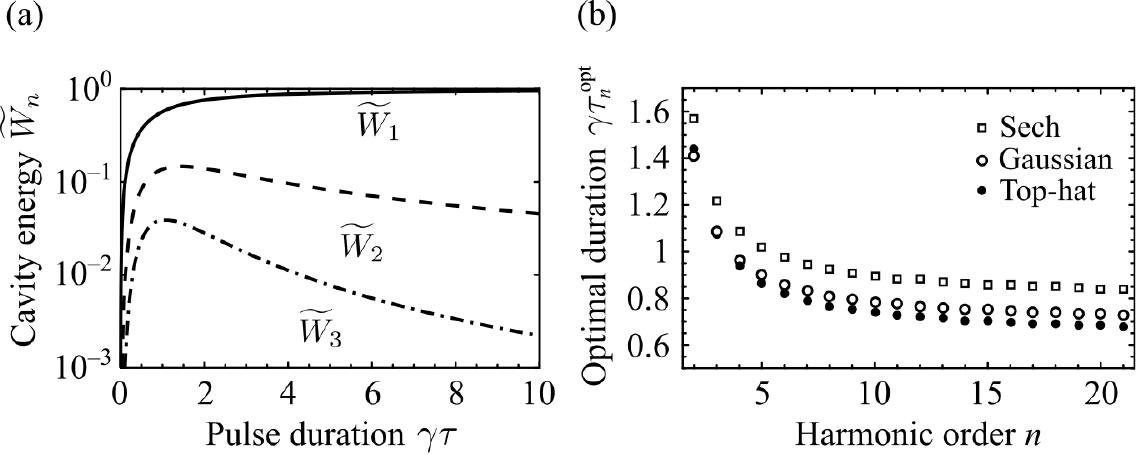}
\caption{Harmonics generation in a nonlinear TVR. (a) Time-integrated normalized energies $\widetilde W_1= \gamma W_1/U_0$ of the fundamental (solid curve), $\widetilde W_2 = \gamma W_2/U_0^2$ of the second-harmonic (dashed curve), and  $\widetilde W_3 = \gamma W_3/U_0^3$ of the third-harmonic (dot-dashed curve) versus normalized pulse duration $\gamma\tau$ under the optimally matched chirp (OMC) excitation by a top-hat pulse. (b) Optimal pulse durations for harmonics orders  $2\leq n\leq 21$ produced by top-hat (dots), Gaussian (circles) and sech-shaped (squares) driving pulses. Note that the optimal pulse duration is approximately equal to the instantaneous fill time $\tau_{\rm res}\equiv\gamma^{-1}$ of the TVR.}\label{fig2}
\end{figure*}

\section{CMT model of a single-mode TVR}\label{sec:CMT}
In the rest of the paper, we assume that a TVR supports a single mode with a time-dependent natural frequency $\omega(t)$, as illustrated in Fig.~\ref{fig1}(a). The TVR has decay time $\tau_{\rm res} \equiv \gamma^{-1}$, and a time-dependent complex amplitude $a(t)$ excited by a free-space excitation field $s(t)$ according to the following CMT equation \cite{Haus1984,Fan2003a}:
\begin{equation}
\dot{a}(t) + [i\omega(t)+\gamma] a(t) = \sqrt{\gamma}s(t),
\label{eq:CMT}
\end{equation}
where the non-radiative (e.g., Ohmic) losses in the resonator are neglected. For simplicity, $\omega(t)$ is assumed to be prescribed and independent of $s(t)$ or $a(t)$, i.e. the nonlinear modifications of the modal frequency due to Kerr effect \cite{Boyd2008} or photo-carrier generation \cite{Preble2007,Tanabe2009} are assumed small. The exact solution of Eq.~(\ref{eq:CMT}), under the initial condition of $a(t=-\infty)=0$, is analytically calculated by transforming into a rotating frame of the resonator's mode and expressing the pulse shape in the form of $s(t)=A(t)e^{-i\psi(t)}$. Here $A(t)$ and $\psi(t)$ are the pulse amplitude and phase, respectively, and the normalized pulse energy is $U_0 = \int^\infty_{-\infty}A^2 dt$. The mode's amplitude is then given by
\begin{equation}\label{eq:CMT-solution}
a(t) = \sqrt{\gamma}e^{-i\phi(t)} \int^t_{-\infty} A(t^\prime) e^{-\gamma(t-t^\prime)} e^{i[\phi(t^\prime)-\psi(t^\prime)]}dt^\prime ,
\end{equation}
where $\phi(t)=\int^t_{-\infty}\omega(t^\prime)dt^\prime$ is the phase advance of the mode.

\section{Optimal temporal format of the driving pulse}\label{sec:OMP}
The mode energy is maximized when the oscillating phase term under the integral is minimized by prescribing
\begin{equation}\label{eq:OMC}
\psi(t)=\int^t_{-\infty}\omega(t^\prime)dt^\prime + \varphi, \mbox{   or   } \dot\psi(t)=\dot\phi(t).
\end{equation}
While the arbitrary geometrical phase $\varphi$ has interesting implications in topological photonics \cite{Lu2014,Khanikaev2017}, it will be omitted in order to stay within the scope of this paper. Equation~(\ref{eq:OMC}), which we refer to as the optimally matched chirp (OMC) condition, provides the recipe for choosing the normalized frequency chirp $\beta$ of the excitation pulse, defined as $\beta \equiv \omega_0^{-2} d^2\psi/dt^2$. For the prescribed modal frequency $\omega(t)$ satisfying the initial condition $\omega(t=0)=\omega_0$, the OMC condition becomes $\beta = \omega_0^{-2} d\omega/dt$. For example, the optimal normalized chirp $\beta$ is constant for a linearly changing modal frequency: if
\begin{equation}\label{eq:wt}
\omega(t) = \omega_0(1 + \alpha\omega_0t),
\end{equation}
then $\phi(t)=\omega_0t + \alpha\omega_0^2 t^2/2$, and Eq.~(\ref{eq:OMC}) yields $\beta=\alpha$.

To illustrate the importance of the optimal chirp, the overall time-integrated energy inside the TVR, $W_1 = \int^\infty_{-\infty}|a(t)|^2dt$, is plotted in Fig.~\ref{fig1}(b) as a function of the input pulse chirp $\beta$ for several values of the damping rate $\gamma$. Here, linearly chirped pulses with $\psi(t) = \omega_0t(1 + \beta\omega_0t/2)$ and Gaussian shapes $A(t)=A_{\rm G}\equiv(2/\pi\tau^2)^{1/4}e^{-t^2/\tau^2}$ are used. The spectral bandwidth of the TVRs is varied from $\gamma=10^{-3}\omega_0$ to $\gamma=10^{-2}\omega_0$  (i.e.  from $Q\equiv\omega_0/2\gamma=500$ to $Q=50$), and the spectral bandwidth swept by the resonant modal frequency over the time duration of the pulse is fixed at $\Delta\omega \approx \alpha\omega_0^2\tau = 0.1\omega_0$ for all cases by choosing $\tau=100/\omega_0$ and $\alpha=10^{-3}$. In other words, in all the cases shown in Fig.~\ref{fig1}(b) the bandwidth sweep $\Delta\omega$ is broader than the TVR's instantaneous bandwidth $\gamma$.

We observe from Fig.~\ref{fig1}(b) that the value of the normalized ${W}_1/U_0$ is maximized for all values of $\gamma$ by satisfying the OMC condition, thereby potentially enhancing any nonlinear interaction between the pulse and the resonator in the presence of a nonlinear material embedded into the resonator cavity. Note that the OMC condition also implies that the resonant modal frequency is swept precisely over the bandwidth of the chirped pulse, i.e. $\Delta\omega_L \equiv \beta\omega_0^2\tau = \Delta\omega$. Therefore, under the OMC condition, the pulse is not transform-limited: $\Delta\omega_L\tau\gg1$. We also observe from Fig.1(b) that, under the OMC condition, the energy enhancement inside the TVR increases with its $Q$-factor. This observation justifies the intuitive approach of using resonators with a narrow linewidth $\gamma$, whose resonant frequency $\omega(t)$ is dynamically swept over a much broader spectral range  $\Delta\omega\gg\gamma$ in order to produce intense broadband harmonic signals. The resulting condition, $\beta\omega_0\tau\gg\gamma/\omega_0$, is satisfied for the three TVRs modeled in Fig.~\ref{fig1}(b).

To investigate the effect of the pulse duration on the total energy pumped into the TVR by the OMC pulse, we calculated the normalized internal TVR energy $\widetilde W_1=\gamma W_1/U_0$ for a top-hat laser pulse $A(t)=A_{\rm TH}$, where $A_{\rm TH}=(2\tau)^{-1/2}$ for $|t|<\tau$. The analytic solution $\widetilde W_1 = 1 + (e^{-2\gamma\tau}-1)/2\gamma\tau$ in Fig.~\ref{fig2}(a) (solid line) versus the normalized pulse duration $\gamma \tau$. The monotonic increase of $\widetilde{W}_1$ with $\tau/\tau_{\rm res}$ (where $\tau_{\rm res}=\gamma^{-1}$ is the instantaneous fill time of the TVR) is expected: pulses that are shorter than $\tau_{\rm res}$ only partially excite the TVR during the time course of their interaction. Note that, for a given chirp rate $\beta = \alpha$ and a given instantaneous $Q$-factor, both the time-integrated energy and bandwidth of the optical field inside the TVR field monotonically increase with the pulse duration.

Therefore, and crucially, the driving pulse does not need to be narrowband to produce strong field enhancement inside a high-$Q$ resonator, as it would be the case for a static resonator. The $W_1\propto Q$ scaling clearly indicates that the combination of a TVR and an optimally chirped excitation signal enables one to resonantly couple a broadband pulse to an evolving resonator with an arbitrarily high instantaneous $Q$-factor. In other words, it is possible to benefit from the resonant energy buildup in a high-$Q$ resonator without sacrificing broadband operation. These conclusions are not altered for the more realistic Gaussian and sech-shaped ($A_{\rm sech}(t) \equiv(2\tau)^{-1} \mbox{sech}(t/\tau)$) OMC pulses, as we discuss below.

\section{Generation of optical harmonics inside a TVR}\label{sec:HHG_TVR}

The concept of efficient cavity excitation by OMC pulses developed in Section~\ref{sec:OMP} is especially relevant for nonlinear optics applications, such  as HG, because the efficiency of the $n$th order process that involves utilizing nonlinear material inside a resonator scales roughly as $|a|^{2n-2}$, i.e. is greatly improved by resonant excitation. Qualitatively, under the undepleted pump approximation and an isotropic nonlinear response, the nonlinear polarization $p^{(n)}$ of the material inside the resonator is  $p^{(n)} = \chi^{(n)}a^n(t)$, where $\chi^{(n)}$ is the nonlinear susceptibility of the resonator material. The emitted $n$th harmonic energy is then proportional to the following quantity:
\begin{equation}
U_n \propto \int|p^{(n)}(t^\prime)|^2dt^\prime = |\chi^{(n)}|^2 W_n(\gamma\tau),
\label{eq:perturb}
\end{equation}
where the duration-dependent $W_n(\gamma\tau) = \int|a(t^\prime)|^{2n} dt^\prime$ is convenient to express as a normalized quantity: $\widetilde{W}_n(\gamma \tau) = \gamma W_n(\gamma\tau)/U_0^n$.

\begin{figure*}\label{fig3}
	\includegraphics[width=0.95\textwidth]{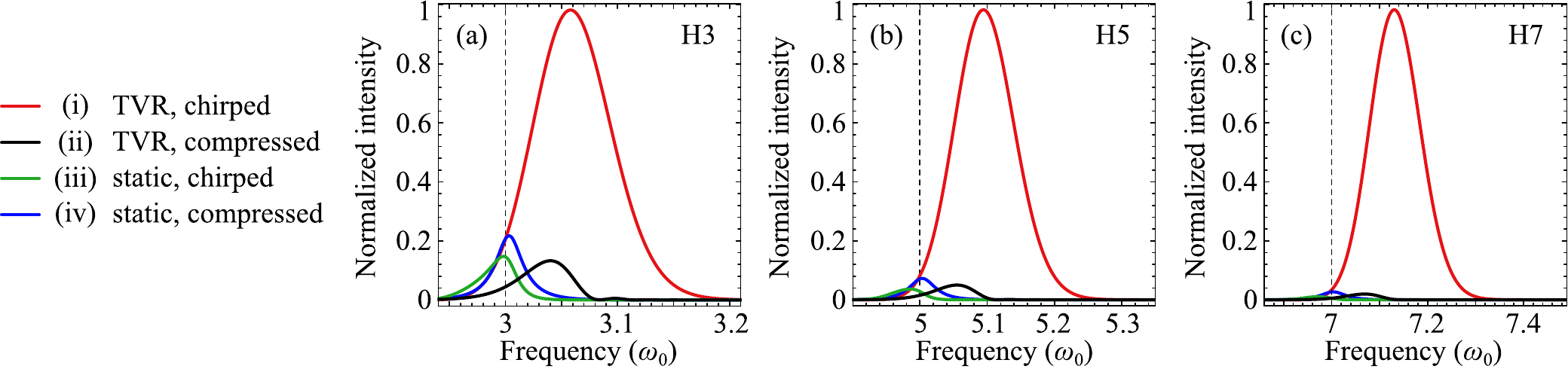}
	\caption{Comparison of harmonics generation using chirped (red and green lines) versus Fourier-transform-limited (black and blue lines) excitation pulses interacting with static (green and blue lines) and time-varying (red and black lines) nonlinear resonators for (a) 3rd, (b) 5th, and (c) 7th harmonics. Instantaneous fill time of all resonators: $\tau_{\rm res}=10^3 \omega_0^{-1}$, compressed pulse duration: $\tau_{\rm comp}=33\omega_0^{-1}$, chirped pulse duration: $\tau_{\rm chirp}=200\omega_0^{-1}$. Normalized sweep rate of the TVR's resonant frequency: $\alpha=1.5\cdot10^{-4}$. The frequency of the mode shifts by $\Delta\omega/\omega_0\approx0.03$ within the chirped pulse. Dashed vertical lines: $\omega_n = n\omega_0$, $n = 3,5,7$.}\label{fig3}
\end{figure*}

The corresponding values of $\widetilde{W}_2$ (dashed line) and $\widetilde{W}_3$ (dot-dashed line) corresponding to the 2nd and 3rd harmonics radiation are plotted in Fig.~\ref{fig2}(a) for different input pulse durations. Remarkably, we observe that, unlike $\widetilde{W}_1$, these quantities are maximized for specific values of the optimal top-hat pulse duration: $\tau_2^{\rm opt}=1.44\gamma^{-1}$ and $\tau_3^{\rm opt}=1.07\gamma^{-1}$. The reason for such a difference between $\widetilde W_1$ and $\widetilde W_n$ ($n\geq2$) is that the nonlinear harmonic conversion is highly sensitive to the peak power, which diminishes as the pulse gets longer. For the 2nd harmonic generation, we can find $\tau_2^{\rm opt}$ from the following analytic dependence of $\widetilde W_2$:
\begin{equation}
\widetilde W_2 (\gamma \tau) = \frac{12 \gamma\tau -11 + 2e^{-6\gamma\tau} -9e^{-4\gamma\tau} + 18e^{-2\gamma\tau} }{24\gamma^2\tau^2}.
\end{equation}

While the analytic expressions for $\widetilde W_n(\gamma \tau)$ are complex even for the top-hat pulses, we have numerically demonstrated the existence and calculated the values of the optimal pulse durations $\tau_n^{\rm opt}$ for $2\leq n\leq21$ that maximize the $n$th harmonic generation. The values of $\tau^{\rm opt}_n/\tau_{\rm res}$ are plotted in Fig.~\ref{fig2}(b) as a function of $n$ for the top-hat (dots),  Gaussian (circles), and sech-shaped (squares) excitation pulses. We observe from Fig.~\ref{fig2}(b) that, regardless of the specific pulse shape, all the optimal pulse durations are close to $\tau_n^{\rm opt}\approx\gamma^{-1}$.  The differences between $\tau_m^{\rm opt}$ and $\tau_l^{\rm opt}$for the $m$th and $l$th order nonlinear processes may be used for adjusting their relative contributions. Notably, the optimal pulse duration does not depend on either the specific functional dependence of $\omega(t)$, nor on the spectral content of the input pulse, as long as the two are matched via the OMC condition given by Eq.~(\ref{eq:OMC}). Therefore, for a given $\tau_n^{\rm opt}$, efficient interaction of the excitation pulse with the resonator can be accomplished for \emph{any} bandwidth of the excitation embedded in $\psi(t)$, as long as the OMC condition is satisfied.

This remarkable ability of TVRs to provide strong field enhancement over a broad bandwidth is illustrated below by comparing the HHG spectra for TVRs (with $\omega(t)$ given by Eq.~(\ref{eq:wt})) versus static resonators (with $\omega(t)=\omega_0$), and for the optimally-chirped pulses versus transform-limited (compressed) pulses with identical power spectra $P_1(\omega)$ defined by Eq.~(\ref{eq:spectrum_n}) below. Note that, in practice, the OMC and transform-limited pulses can be converted into each other by propagating through a dispersive material, or reflecting off a pair of gratings \cite{Strickland1985,Weiner2011}.

The normalized  $n$'th harmonic power spectra, defined as
\begin{equation}\label{eq:spectrum_n}
P_n(\omega)\propto \left|\int a^n(t^\prime)e^{i\omega t^\prime}dt^\prime \right|^2,
\end{equation}
are plotted in Figs.~\ref{fig3}(a-c) for the $n=3,5,7$ harmonics, respectively. For the resonator and pulse parameters given in the caption, the bandwidth of the incident pulses is much wider than the instantaneous bandwidth of the resonators: $\Delta \omega_L/\omega_0 \approx 0.07 \gg Q^{-1}$, where $Q=200$ for both static and time-varying resonators. For example, the normalized third harmonic (H3) spectra shown in Fig.~\ref{fig3}(a) correspond to the following four cases: (i) a TVR driven by an OMC pulse (red curve) with duration $\tau_{\rm chirp}=200\omega_0^{-1}$; (ii) the same TVR excited by a transform-limited pulse with the same spectrum and fluence, but shorter duration $\tau_{\rm comp}=33\omega_0^{-1}$ (black curve); (iii) a static resonator excited by the OMC pulse (green curve); and (iv) the same static resonator excited by the transform-limited pulse (blue curve). We note that for the optimal case (i), the following relation holds for the $n$-th harmonic spectra $\int P_n^{\rm OMC}(\omega,\tau_{\rm chirp})$: $\int P_n^{\rm OMC} d\omega = \left( U_0^{n}/\gamma \right) \widetilde{W}_n(\gamma \tau_n^{\rm chirp})$, where, as an example, $\widetilde{W}_3$ is plotted in Fig.~\ref{fig2}(a).

The limitations of using static resonators in cases (iii-iv) are apparent from Fig.~\ref{fig3}(a): they produce the weakest H3 spectra. Even though the peak intensity of the transform-limited pulses is the highest, the resulting H3 spectrum in case (iv) is only slightly stronger than that for the chirped case (iii). The transform-limited pulse is also inefficient in producing the third harmonic when interacting with a TVR (case (ii)), even though a somewhat broader spectrum is obtained in comparison with the cases (iii) and (iv). Note that the observed spectral blue-shifting in case (ii) is a manifestation of the phenomenon of photon acceleration (PA) previously observed in plasmas and dielectric resonators \cite{Wilks1989,Siders1996,Shcherbakov2019}. The strongest and broadest H3 spectrum is produced when the nonlinear TVR interacts with the OMC pulse. The resulting spectrum is significantly blue-shifted due to PA, and broadened: it covers the largest spectral bandwidth of the four cases.

The advantage of combining a TVR with an optimally chirped pulse is even more apparent for HG with $n>3$. The calculations for the spectra of the fifth (H5) and seventh (H7) harmonics are presented in Fig.~\ref{fig3}(b,c) for the above cases (i-iv). By comparing case (i) (red lines) to cases (ii-iv), we observe that the enhancement in both the total energy of the harmonics (areas under the curves) and in  their spectral bandwidth dramatically increases with increasing harmonic number $n$.  Our findings strongly suggest that a judicious choice of the signal chirp and of the resonance frequency variation can intensify the HG output of the TVR. Note that both ingredients -- the time-dependent nature of the resonator and the engineered pulse chirp -- must be present in order to achieve ultra-high efficiencies that are illustrated by case (i) in Figs.~\ref{fig3}(a-c).

\begin{figure*}[t!]
	\includegraphics[width=0.6	\textwidth]{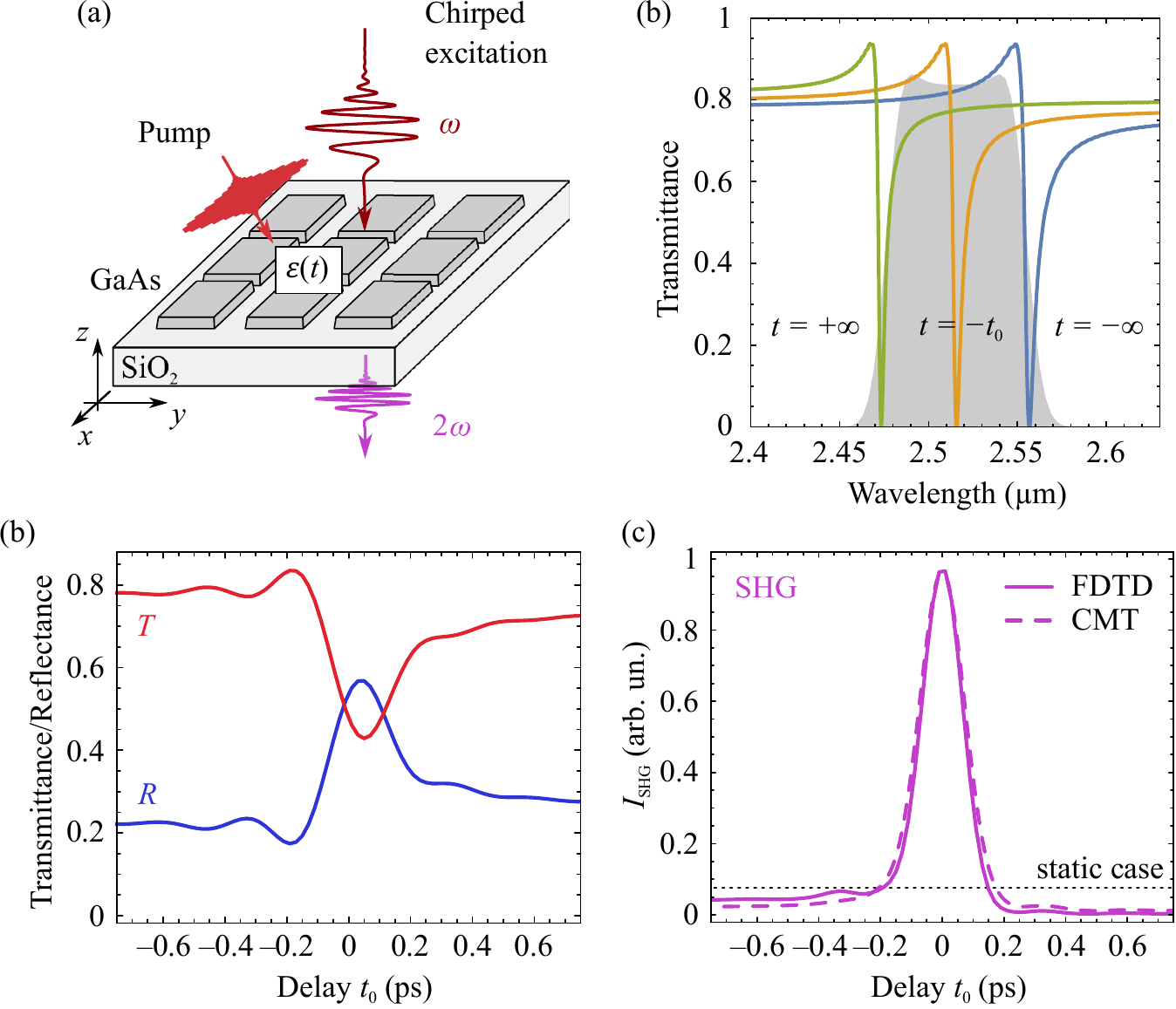}\caption{\label{fig5}
	Implementation of a TVR using a semiconductor metasurface illuminated by a laser pulse (pump). (a) Schematic: a high-$Q$ GaAs-based time-variant metasurface (TVM). The time dependence of the dielectric permittivity $\varepsilon(t)$ is caused by the free carriers produced by the pump. Second harmonic is generated by the chirped OMC excitation pulse delayed in time by $t_0$ from the pump. The excitation pulse has a duration of 500~fs and a carrier wavelength of 2.514~$\mu$m. (b) Transmittance/reflectance spectra for static metasurfaces with $\varepsilon(t) \equiv \varepsilon(-\infty)$ (blue), $\varepsilon(-t_0)$ (orange) and $\varepsilon(\infty)$ (green). (c) Time-integrated transmittance and reflectance of the metasurface as a function of $t_0$. (c) Second harmonic generation as a function of $t_0$, calculated numerically by an FDTD simulation (solid line) and from the CMT (dashed line). Dotted line: SHG for a static metasurface with the resonance centered at $\lambda_R^{(s)} = 2.514\mu$m (i.e., at the carrier wavelength of the pulse). Laser parameters: durations of the pump and excitation pulses $\sigma= 500$~fs, the bandwidth of the excitation pulse $\Delta \omega_L = 34$THz. Metasurface parameters: GaAs bars' sizes $L_x \times L_y\times L_z =$ $1.3\mu$m $\times 0.6 \mu$m $\times 0.2 \mu $m, periodicities $P_x = 1.56~\mu$m and $P_y = 1.2~\mu$m.
	}
\end{figure*}

\section{A Physical Realization of a TVR: A Time-Variant Nonlinear Metasurface}\label{sec:TVM}

As a specific realization of a TVR, we are envisioning a high-$Q$ nonlinear metasurface comprised of a semiconductor material with time-dependent dielectric permittivity $\varepsilon(t)$. To support our findings and to provide a specific blueprint for a nonlinear TVR, below we describe such a time-varying metasurface (TVM). We have performed full-wave electromagnetic simulations using a finite-difference time-domain (FDTD) solver (Lumerical FDTD Solutions) on a metasurface shown in Fig.~\ref{fig5}(a). Metasurfaces have previously shown promise for making efficient nonlinear resonators at the nanoscale \cite{Shcherbakov2014,Yang2015a,Vampa2017,Sivis2017,Shcherbakov2019,Shaltout2015}. The metasurface under study and its physical dimensions are shown in Fig.~\ref{fig5}(a). It was designed to exhibit resonant reflection (transmission) peak (dip) at the permittivity-dependent resonant frequency $\omega_{R}^{(s)} \equiv \omega(\varepsilon_{-\infty})$ corresponding to the {\it static} resonant wavelength $\lambda_R^{(s)} \equiv 2\pi c/\omega_R^{(s)} \approx 2.555~\mu$m.  The baseline (non-resonant) reflection and transmission coefficients were found to be $R_0 \approx 25\%$ and $T_0 \approx 75\%$, respectively. The real-valued (i.e. lossless) time-dependent dielectric permittivity of the metasurface material was modeled as follows:
\begin{equation}
\varepsilon(t) = \varepsilon_{-\infty} + \frac{\Delta\varepsilon}{2}\left[\mbox{erf}\left(\frac{t+t_0}{\sigma}\right) + 1\right],
\label{eq:eps}
\end{equation}
where $\varepsilon_{-\infty}=11.02$ corresponds to GaAs in the near-infrared. We assumed that $\Delta \varepsilon =-2.5$ (i.e. the reduction of the refractive index by approximately $12\%$) is caused by free carrier generation. The time scale of the refractive index variation, and, therefore, of the resulting resonant frequency change, is assumed to be $\sigma = 500$~fs. A specific physical realization of such time dependence of $\varepsilon(t)$ could involve, for example, electron-hole pairs generation in a semiconductor by a Gaussian laser pulse with duration $\sigma$ advanced by the time interval $t_0$ with respect to the excitation pulse. By assuming the material to be transparent, we have neglected any reduction of the quality factor of the metasurface due to the presence of free carriers in the material.

In order to relate the resonant frequency $\omega(t)$ from Eq.(\ref{eq:CMT}) to dielectric permittivity $\varepsilon(t)$ given by Eq.(\ref{eq:eps}), we calculated the passive transmittance spectra of the metasurface as a function of $\varepsilon$ in the range from $\varepsilon_{-\infty}$ to $\varepsilon_{-\infty}+\Delta\varepsilon$. The resulting $\omega(\varepsilon)$ dependence showed to be linear in $\delta \varepsilon \equiv \varepsilon_{-\infty} - \varepsilon$: $\omega(\varepsilon) = \omega_0 + k \delta \varepsilon$,  where $\omega_0=737$~THz and $k =10.2$~THz.
The excitation field is chosen as $E(t) = A(t)e^{i\psi(t)}$, where $A(t) = E_0\exp\left(-t^2/2\sigma^2\right)$, $E_0$ is real-valued. The phase term $\psi(t)$ is calculated according to the OMC condition given by Eq.(\ref{eq:OMC}) for $t_0=0$:
\begin{equation}
\psi(t) =
\frac{\Delta\varepsilon k}{2}
\left[
\frac{\sigma e^{-\frac{t^2} {\sigma^2} } }  {\sqrt{\pi}} +
t\mbox{erf}\left(\frac{t}{\sigma}\right) +
t
\right]
+ \omega_0 t.
\label{eq:psi}
\end{equation}
The full-width at half maximum (fwhm) intensity bandwidth of the excitation pulse $\Delta\omega_{\rm L}=34$~THz, which was chosen to approximately match the change of the resonant frequency of the metasurface due to free-carriers generation, is assumed to be much larger than the instantaneous width $\Delta\omega_{\rm R}=2.2$~THz of the metasurface resonance. Figure~\ref{fig5}(b) shows the total (time-integrated) transmitted and reflected fluence as a function of the time delay $t_0$. We adopt the convention of $t_0>0$ corresponding to the free carrier generating pump arriving before the peak of the mid-infrared probe. We observe that for large values of $|\tau_0|\gg \sigma$ the transmission/reflection curves plotted as a function of $t_0$ are essentially flat, corresponding to the non-resonant (baseline) transmission (reflection) coefficients $T_0$ ($R_0$). This is because most of the incident pulse experiences an essentially time-invariant resonant metasurface. Therefore, only a small portion of the broadband pulse experiences resonant transmission/reflection. The situation dramatically changes for small time delays $|\tau_0| < \sigma$: both the transmittance and reflectance curves exhibit sub-picosecond features as a function of $t_0$. Prominent transmission dip and reflection peak are observed in Figure~\ref{fig5}(b) around the $t_0=0$ value of the time delay.
The reflection enhancement observed around $t_0=0$ is caused by the PA of light trapped inside the TVM~\cite{Shcherbakov2019}: when the frequency of the incoming photons matches that of the continuously accelerated photons, the intensity of light inside the TVM increases, and so does the reflectivity. This effect can be interpreted as constructive interference between different portions of the pulse representing different portion of the spectrum. The above-mentioned frequency matching occurs when the frequency chirp of the incident pulse matches that of the TVM's resonant frequency $\omega(t) \equiv \omega(\varepsilon(t))$.

The resulting energy concentration inside the TVM is accompanied by significant enhancement of all nonlinear processes, including the second harmonic generation (SHG) from a metasurface comprised of a material with a finite $\chi_2$ coefficient (GaAs, in the considered case). To model the SHG inside the structure using the FDTD code from Lumerical, an additional term proportional to the electric field was added to the polarizability of the simulated material; this created an effective source of harmonic radiation inside the metasurface. For simplicity, we neglected the tensor structure of GaAs nonlinear susceptibility and assumed isotropic response. The far-field radiation produced by this source was then integrated over time as a function of time delay $t_0$. The result of this calculation, plotted in Fig.~\ref{fig5}(c) as a solid line, shows more than an order-of-magnitude enhancement in the emitted SHG at $t_0=0$ with respect to the non-resonant value (dotted horizontal line) obtained for a static metasurface resonating at $\lambda_R^{(s)}$.

The dashed line in Fig.~\ref{fig5}(c) shows the results of the CMT modeling (see Section~\ref{sec:CMT}), where the parameters from Eqs.~(\ref{eq:eps},\ref{eq:psi}) were used. The radiative decay rate entering Eq.(\ref{eq:CMT}) was chosen to capture the actual $Q$-factor of the metasurface. The solid and dashed curves were normalized so that their maximum values at $t_0=0$ were the same. The overall agreement of the FDTD and CMT results serves as a verification step for our theory and suggests experimental feasibility of using the chirp-matching schemes to increase the efficiency of nonlinear light-matter interactions in TVMs and TVRs.

\section{Applications of broadband harmonics generation}
\begin{figure}
	\includegraphics[width=0.9\columnwidth]{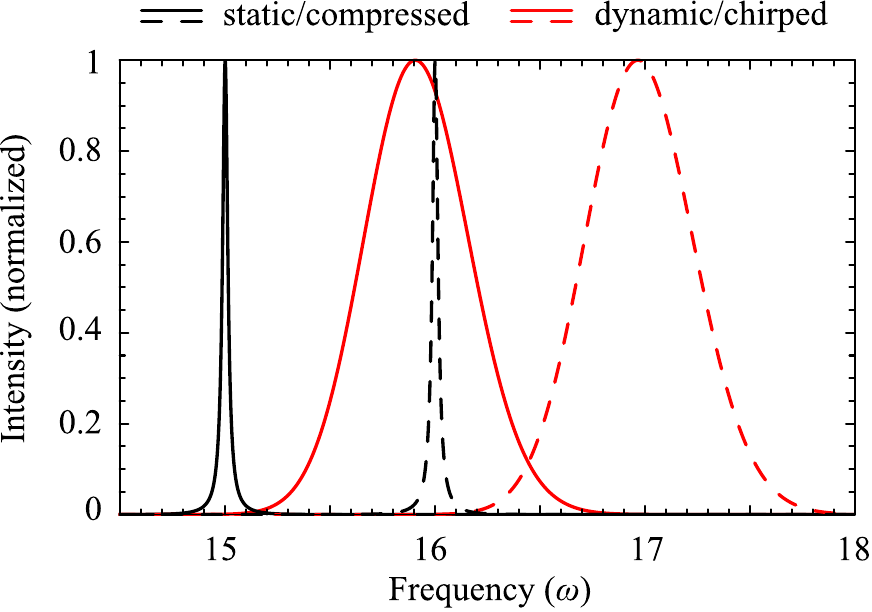}
	\caption{\label{fig4} Overlapping harmonics from TVRs. Spectra of the 15th (solid lines) and 16th (dashed lines) optical harmonics generated by a static resonator and a compressed pulse (black) and a TVR and an OMC-pulse (red), revealing the spectral overlap. The spectra are normalized by unity for clarity.
	}
\end{figure}

High harmonic generation (HHG) from structured surfaces~\cite{Han2016,Sivis2017,Vampa2017,Liu2018f,Ghimire2019} leads to a  variety of exciting applications, including all-optical probing of the electronic band structure~\cite{Anin2017} and the generation of high angular momentum UV beams~\cite{Gariepy2014,Li2017c} for high-resolution lithography~\cite{Wagner2010} and imaging~\cite{Seaberg2011}. Moreover, the utilization of mid-infrared (sub-bandgap) laser pulses can enable efficient HHG from solids through  processes similar to those in gases in the strong field ionization (SFI)  regime~\cite{Ghimire2011,Vampa2014a,Vampa2015,Wolter2015}. A remarkable feature of the SFI (i.e. small Keldysh parameter) regime is that the HHG spectra saturate for large harmonic numbers $N$~\cite{You2017}: $P_{N}(\omega_N) \approx P_{N+1}(\omega_{N+1})$, where $\omega_n = n\omega_0$. If such harmonics are produced inside a high-$Q$ static resonator by a narrow-band laser pulse with frequency $\omega_0$, then the adjacent $\omega_{N}$'th and $\omega_{N+1}$-st harmonics do non-overlap as long as their fractional spectral widths $\Delta \omega_N/\omega_N \approx Q^{-1} \ll 1/N$. The lack of the harmonics overlap is shown in Fig.~\ref{fig4} for $N=15,16$ (black lines) and $Q=10^3$. Therefore, taking advantage of the high-$Q$ (and field-enhancing, see Fig.~\ref{fig1}) resonator for the purpose of ultra-efficient generation of attosecond UV laser pulses is very challenging because the latter requires continuous spectral overlap between multiple harmonics.

As shown in Fig.~\ref{fig4} (red lines), this challenge is overcome by an OMC pulse with $\Delta \omega_L/\omega_0 \approx 0.12$ ($\tau_{\rm comp}=20\omega_0^{-1}$) interacting with a TVR.
For this simulation we have used the laser and TVR parameters similar to the ones used in Fig.~\ref{fig3}: $\tau_{\rm chirp}=10^3\omega_0^{-1}$, and $\alpha = \beta = 10^{-4}$. The two adjacent harmonics have a considerable overlap because of the spectral broadening of both. We propose that the requisite sweep of the TVR's resonant frequency $\omega(t)$ within the laser pulse duration can be achieved via photoinjection of free carriers by ultra-short laser pulses \cite{Preble2012,Notomi2007a}. A promising platform for such a TVR is a semiconductor metasurface interacting with temporally overlapping short-wavelength ($\lambda_{\rm pump}$) carrier-generating laser pump and a long-wavelength ($\lambda_{\rm probe} \gg \lambda_{\rm pump}$) harmonics-generating laser probe~\cite{Shcherbakov2019}. By choosing $\lambda_{\rm probe}$ and $\lambda_{\rm probe}$ in the mid-infrared and visible/near-infrared spectral ranges, respectively, one can ensure that (i) single-photon free-carrier generation by the pump dominates over that by the probe despite the probe's higher intensity; (ii) the HHG occurs in the SFI regime at modest intensities because the $\lambda^{-1}$ scaling of the Keldysh parameter favor the mid-IR probe \cite{Vampa2014a}.

Using GaAs as a TVR material, we note that a free-carrier density $\rho \approx 3 \cdot 10^{18}$~cm$^{-3}$\cite{Bennett1990} produced by a pump laser with $\lambda_{\rm pump} \sim 0.8$~$\mu$m can induce the required frequency shifts of a TVR tuned to $\lambda_{\rm probe} \sim 4$~$\mu$m. Because intensity enhancements of order $I_{\rm hot} \sim Q I_0$ can be expected inside the hot spots of a metasurface, we anticipate that even modest incident intensities $I_0 \sim 10^{10}$~W/cm$^2$ can produce spectrally overlapping harmonics with $N\sim 15$ or higher~\cite{You2017}. Even though the emitting volume of these resonators is severely restricted by single-photon absorption of HHG photons \cite{Ghimire2012}, the overall HHG yield will be multifold with respect to the unstructured material. These estimates suggest that employing the concept described in this paper -- of a high-$Q$ time-varying resonator interacting with broadband optimally-chirped laser pulses -- could provide an important platform for realizing numerous applications of attosecond UV pulses. Although Eq.(\ref{eq:perturb}) describes perturbative nonlinearities that are not always applicable to HHG, we think that the chirp-matching approach can still be successfully applied to HHG in resonators, since HHG strongly depends on the local fields within a TVR. We think that further studies that involve full quantum-mechanical description of HHG \cite{Vampa2014a} are required to verify these exciting findings.
We envision that the chirp-matching mechanism of producing strong spectral overlap between high laser harmonics produced in the SFI regime will also contribute to generating novel states of light, e.g., of beams with mixed high orbital angular momenta~\cite{Gariepy2014}.

To conclude, we have analyzed a coupled-mode theory of a time-varying resonator (TVR) under pulsed excitation. From the obtained exact soultion, we have found the optimal excitation pulse format (chirp and duration) for any prescribed time evolution of the resonator. Both linear and nonlinear responses were used as optimization targets. We have shown that the bandwidth of the light-resonator interaction can be decoupled from the instantaneous spectral selectivity of the TVR. Because the latter determines the field enhancement inside the resonator, and, therefore, the efficiency of any nonlinear optical process, TVRs can be used to overcome the efficiency-bandwidth tradeoff that is inherent in time-invariant resonators. Finally, we propose a specific TVR platform -- a high-$Q$ time-variant semiconductor metasurface -- that enables efficient generation of broadband nonlinear signals. This platform can be realistically implemented in modern high-$Q$ systems by means of photoinduced free carrier generation. Our results pave the way to efficient broadband generation of optical harmonics, which can be further employed to produce attosecond pulses using moderate-power lasers. Given the general nature of our theory, it can be applicable to other types of driven resonators, including radio \cite{Liu2018g}, mechanical, and acoustic ones.

This work was supported by the Office of Naval Research (ONR) Grant N00014-17-1-2161 and Russian Science Foundation (grant 18-12-00475, FDTD simulations).


\begin{thebibliography}{48}%
\makeatletter
\providecommand \@ifxundefined [1]{%
 \@ifx{#1\undefined}
}%
\providecommand \@ifnum [1]{%
 \ifnum #1\expandafter \@firstoftwo
 \else \expandafter \@secondoftwo
 \fi
}%
\providecommand \@ifx [1]{%
 \ifx #1\expandafter \@firstoftwo
 \else \expandafter \@secondoftwo
 \fi
}%
\providecommand \natexlab [1]{#1}%
\providecommand \enquote  [1]{``#1''}%
\providecommand \bibnamefont  [1]{#1}%
\providecommand \bibfnamefont [1]{#1}%
\providecommand \citenamefont [1]{#1}%
\providecommand \href@noop [0]{\@secondoftwo}%
\providecommand \href [0]{\begingroup \@sanitize@url \@href}%
\providecommand \@href[1]{\@@startlink{#1}\@@href}%
\providecommand \@@href[1]{\endgroup#1\@@endlink}%
\providecommand \@sanitize@url [0]{\catcode `\\12\catcode `\$12\catcode
  `\&12\catcode `\#12\catcode `\^12\catcode `\_12\catcode `\%12\relax}%
\providecommand \@@startlink[1]{}%
\providecommand \@@endlink[0]{}%
\providecommand \url  [0]{\begingroup\@sanitize@url \@url }%
\providecommand \@url [1]{\endgroup\@href {#1}{\urlprefix }}%
\providecommand \urlprefix  [0]{URL }%
\providecommand \Eprint [0]{\href }%
\providecommand \doibase [0]{http://dx.doi.org/}%
\providecommand \selectlanguage [0]{\@gobble}%
\providecommand \bibinfo  [0]{\@secondoftwo}%
\providecommand \bibfield  [0]{\@secondoftwo}%
\providecommand \translation [1]{[#1]}%
\providecommand \BibitemOpen [0]{}%
\providecommand \bibitemStop [0]{}%
\providecommand \bibitemNoStop [0]{.\EOS\space}%
\providecommand \EOS [0]{\spacefactor3000\relax}%
\providecommand \BibitemShut  [1]{\csname bibitem#1\endcsname}%
\let\auto@bib@innerbib\@empty
\bibitem [{\citenamefont {Luchansky}\ and\ \citenamefont
  {Bailey}(2012)}]{Luchansky2012}%
  \BibitemOpen
  \bibfield  {author} {\bibinfo {author} {\bibfnamefont {M.~S.}\ \bibnamefont
  {Luchansky}}\ and\ \bibinfo {author} {\bibfnamefont {R.~C.}\ \bibnamefont
  {Bailey}},\ }\href {\doibase 10.1021/ac2029024.High-Q} {\bibfield  {journal}
  {\bibinfo  {journal} {Anal. Chem.}\ }\textbf {\bibinfo {volume} {84}},\
  \bibinfo {pages} {793} (\bibinfo {year} {2012})}\BibitemShut {NoStop}%
\bibitem [{\citenamefont {Reiserer}\ and\ \citenamefont
  {Rempe}(2015)}]{Reiserer2015}%
  \BibitemOpen
  \bibfield  {author} {\bibinfo {author} {\bibfnamefont {A.}~\bibnamefont
  {Reiserer}}\ and\ \bibinfo {author} {\bibfnamefont {G.}~\bibnamefont
  {Rempe}},\ }\href {\doibase 10.1103/RevModPhys.87.1379} {\bibfield  {journal}
  {\bibinfo  {journal} {Rev. Mod. Phys.}\ }\textbf {\bibinfo {volume} {87}},\
  \bibinfo {pages} {1379} (\bibinfo {year} {2015})}\BibitemShut {NoStop}%
\bibitem [{\citenamefont {{LIGO Collaboration}}(2013)}]{Aasi2013}%
  \BibitemOpen
  \bibfield  {author} {\bibinfo {author} {\bibnamefont {{LIGO
  Collaboration}}},\ }\href {\doibase 10.1038/nphoton.2013.177} {\bibfield
  {journal} {\bibinfo  {journal} {Nature Photon.}\ }\textbf {\bibinfo {volume}
  {7}},\ \bibinfo {pages} {613} (\bibinfo {year} {2013})}\BibitemShut {NoStop}%
\bibitem [{\citenamefont {Shcherbakov}\ \emph {et~al.}(2014)\citenamefont
  {Shcherbakov}, \citenamefont {Neshev}, \citenamefont {Hopkins}, \citenamefont
  {Shorokhov}, \citenamefont {Staude}, \citenamefont {Melik-Gaykazyan},
  \citenamefont {Decker}, \citenamefont {Ezhov}, \citenamefont
  {Miroshnichenko}, \citenamefont {Brener}, \citenamefont {Fedyanin},\ and\
  \citenamefont {Kivshar}}]{Shcherbakov2014}%
  \BibitemOpen
  \bibfield  {author} {\bibinfo {author} {\bibfnamefont {M.~R.}\ \bibnamefont
  {Shcherbakov}}, \bibinfo {author} {\bibfnamefont {D.~N.}\ \bibnamefont
  {Neshev}}, \bibinfo {author} {\bibfnamefont {B.}~\bibnamefont {Hopkins}},
  \bibinfo {author} {\bibfnamefont {A.~S.}\ \bibnamefont {Shorokhov}}, \bibinfo
  {author} {\bibfnamefont {I.}~\bibnamefont {Staude}}, \bibinfo {author}
  {\bibfnamefont {E.~V.}\ \bibnamefont {Melik-Gaykazyan}}, \bibinfo {author}
  {\bibfnamefont {M.}~\bibnamefont {Decker}}, \bibinfo {author} {\bibfnamefont
  {A.~A.}\ \bibnamefont {Ezhov}}, \bibinfo {author} {\bibfnamefont {A.~E.}\
  \bibnamefont {Miroshnichenko}}, \bibinfo {author} {\bibfnamefont
  {I.}~\bibnamefont {Brener}}, \bibinfo {author} {\bibfnamefont {A.~A.}\
  \bibnamefont {Fedyanin}}, \ and\ \bibinfo {author} {\bibfnamefont {Y.~S.}\
  \bibnamefont {Kivshar}},\ }\href@noop {} {\bibfield  {journal} {\bibinfo
  {journal} {Nano Lett.}\ }\textbf {\bibinfo {volume} {14}},\ \bibinfo {pages}
  {6488} (\bibinfo {year} {2014})}\BibitemShut {NoStop}%
\bibitem [{\citenamefont {Yang}\ \emph {et~al.}(2015)\citenamefont {Yang},
  \citenamefont {Wang}, \citenamefont {Boulesbaa}, \citenamefont {Kravchenko},
  \citenamefont {Briggs}, \citenamefont {Puretzky}, \citenamefont {Geohegan},\
  and\ \citenamefont {Valentine}}]{Yang2015a}%
  \BibitemOpen
  \bibfield  {author} {\bibinfo {author} {\bibfnamefont {Y.}~\bibnamefont
  {Yang}}, \bibinfo {author} {\bibfnamefont {W.}~\bibnamefont {Wang}}, \bibinfo
  {author} {\bibfnamefont {A.}~\bibnamefont {Boulesbaa}}, \bibinfo {author}
  {\bibfnamefont {I.~I.}\ \bibnamefont {Kravchenko}}, \bibinfo {author}
  {\bibfnamefont {D.~P.}\ \bibnamefont {Briggs}}, \bibinfo {author}
  {\bibfnamefont {A.}~\bibnamefont {Puretzky}}, \bibinfo {author}
  {\bibfnamefont {D.}~\bibnamefont {Geohegan}}, \ and\ \bibinfo {author}
  {\bibfnamefont {J.}~\bibnamefont {Valentine}},\ }\href {\doibase
  10.1021/acs.nanolett.5b02802} {\bibfield  {journal} {\bibinfo  {journal}
  {Nano Lett.}\ }\textbf {\bibinfo {volume} {15}},\ \bibinfo {pages} {7388}
  (\bibinfo {year} {2015})}\BibitemShut {NoStop}%
\bibitem [{\citenamefont {Vabishchevich}\ \emph {et~al.}(2018)\citenamefont
  {Vabishchevich}, \citenamefont {Liu}, \citenamefont {Sinclair}, \citenamefont
  {Keeler}, \citenamefont {Peake},\ and\ \citenamefont
  {Brener}}]{Vabishchevich2018}%
  \BibitemOpen
  \bibfield  {author} {\bibinfo {author} {\bibfnamefont {P.~P.}\ \bibnamefont
  {Vabishchevich}}, \bibinfo {author} {\bibfnamefont {S.}~\bibnamefont {Liu}},
  \bibinfo {author} {\bibfnamefont {M.~B.}\ \bibnamefont {Sinclair}}, \bibinfo
  {author} {\bibfnamefont {G.~A.}\ \bibnamefont {Keeler}}, \bibinfo {author}
  {\bibfnamefont {G.~M.}\ \bibnamefont {Peake}}, \ and\ \bibinfo {author}
  {\bibfnamefont {I.}~\bibnamefont {Brener}},\ }\href {\doibase
  10.1021/acsphotonics.7b01478} {\bibfield  {journal} {\bibinfo  {journal} {ACS
  Photonics}\ }\textbf {\bibinfo {volume} {5}},\ \bibinfo {pages} {1685}
  (\bibinfo {year} {2018})}\BibitemShut {NoStop}%
\bibitem [{\citenamefont {Vampa}\ \emph {et~al.}(2017)\citenamefont {Vampa},
  \citenamefont {Ghamsari}, \citenamefont {{Siadat Mousavi}}, \citenamefont
  {Hammond}, \citenamefont {Olivieri}, \citenamefont {Lisicka-Skrek},
  \citenamefont {Naumov}, \citenamefont {Villeneuve}, \citenamefont {Staudte},
  \citenamefont {Berini},\ and\ \citenamefont {Corkum}}]{Vampa2017}%
  \BibitemOpen
  \bibfield  {author} {\bibinfo {author} {\bibfnamefont {G.}~\bibnamefont
  {Vampa}}, \bibinfo {author} {\bibfnamefont {B.~G.}\ \bibnamefont {Ghamsari}},
  \bibinfo {author} {\bibfnamefont {S.}~\bibnamefont {{Siadat Mousavi}}},
  \bibinfo {author} {\bibfnamefont {T.~J.}\ \bibnamefont {Hammond}}, \bibinfo
  {author} {\bibfnamefont {A.}~\bibnamefont {Olivieri}}, \bibinfo {author}
  {\bibfnamefont {E.}~\bibnamefont {Lisicka-Skrek}}, \bibinfo {author}
  {\bibfnamefont {A.~Y.}\ \bibnamefont {Naumov}}, \bibinfo {author}
  {\bibfnamefont {D.~M.}\ \bibnamefont {Villeneuve}}, \bibinfo {author}
  {\bibfnamefont {A.}~\bibnamefont {Staudte}}, \bibinfo {author} {\bibfnamefont
  {P.}~\bibnamefont {Berini}}, \ and\ \bibinfo {author} {\bibfnamefont {P.~B.}\
  \bibnamefont {Corkum}},\ }\href {\doibase 10.1038/nphys4087} {\bibfield
  {journal} {\bibinfo  {journal} {Nat. Phys.}\ }\textbf {\bibinfo {volume}
  {13}},\ \bibinfo {pages} {659} (\bibinfo {year} {2017})}\BibitemShut
  {NoStop}%
\bibitem [{\citenamefont {Sivis}\ \emph {et~al.}(2017)\citenamefont {Sivis},
	\citenamefont {Taucer}, \citenamefont {Vampa}, \citenamefont {Johnston},
	\citenamefont {Staudte}, \citenamefont {Naumov}, \citenamefont {Villeneuve},
	\citenamefont {Ropers},\ and\ \citenamefont {Corkum}}]{Sivis2017}%
\BibitemOpen
\bibfield  {author} {\bibinfo {author} {\bibfnamefont {M.}~\bibnamefont
		{Sivis}}, \bibinfo {author} {\bibfnamefont {M.}~\bibnamefont {Taucer}},
	\bibinfo {author} {\bibfnamefont {G.}~\bibnamefont {Vampa}}, \bibinfo
	{author} {\bibfnamefont {K.}~\bibnamefont {Johnston}}, \bibinfo {author}
	{\bibfnamefont {A.}~\bibnamefont {Staudte}}, \bibinfo {author} {\bibfnamefont
		{A.~Y.}\ \bibnamefont {Naumov}}, \bibinfo {author} {\bibfnamefont {D.~M.}\
		\bibnamefont {Villeneuve}}, \bibinfo {author} {\bibfnamefont
		{C.}~\bibnamefont {Ropers}}, \ and\ \bibinfo {author} {\bibfnamefont {P.~B.}\
		\bibnamefont {Corkum}},\ }\href@noop {} {\bibfield  {journal} {\bibinfo
		{journal} {Science}\ }\textbf {\bibinfo {volume} {306}},\ \bibinfo {pages}
	{303} (\bibinfo {year} {2017})}\BibitemShut {NoStop}%
\bibitem [{\citenamefont {Tsakmakidis}\ \emph {et~al.}(2017)\citenamefont
  {Tsakmakidis}, \citenamefont {Shen}, \citenamefont {Schulz}, \citenamefont
  {Zheng}, \citenamefont {Upham}, \citenamefont {Deng}, \citenamefont {Altug},
  \citenamefont {Vakakis},\ and\ \citenamefont {Boyd}}]{Tsakmakidis2017}%
  \BibitemOpen
  \bibfield  {author} {\bibinfo {author} {\bibfnamefont {K.~L.}\ \bibnamefont
  {Tsakmakidis}}, \bibinfo {author} {\bibfnamefont {L.}~\bibnamefont {Shen}},
  \bibinfo {author} {\bibfnamefont {S.~A.}\ \bibnamefont {Schulz}}, \bibinfo
  {author} {\bibfnamefont {X.}~\bibnamefont {Zheng}}, \bibinfo {author}
  {\bibfnamefont {J.}~\bibnamefont {Upham}}, \bibinfo {author} {\bibfnamefont
  {X.}~\bibnamefont {Deng}}, \bibinfo {author} {\bibfnamefont {H.}~\bibnamefont
  {Altug}}, \bibinfo {author} {\bibfnamefont {A.~F.}\ \bibnamefont {Vakakis}},
  \ and\ \bibinfo {author} {\bibfnamefont {R.~W.}\ \bibnamefont {Boyd}},\
  }\href {\doibase 10.1126/science.aam6662} {\bibfield  {journal} {\bibinfo
  {journal} {Science}\ }\textbf {\bibinfo {volume} {356}},\ \bibinfo {pages}
  {1260} (\bibinfo {year} {2017})}\BibitemShut {NoStop}%
\bibitem [{\citenamefont {Mann}\ \emph {et~al.}(2019)\citenamefont {Mann},
  \citenamefont {Sounas},\ and\ \citenamefont {Alu}}]{Mann2019}%
  \BibitemOpen
  \bibfield  {author} {\bibinfo {author} {\bibfnamefont {S.~A.}\ \bibnamefont
  {Mann}}, \bibinfo {author} {\bibfnamefont {D.~L.}\ \bibnamefont {Sounas}}, \
  and\ \bibinfo {author} {\bibfnamefont {A.}~\bibnamefont {Alu}},\ }\href@noop
  {} {\bibfield  {journal} {\bibinfo  {journal} {Optica}\ }\textbf {\bibinfo
  {volume} {6}},\ \bibinfo {pages} {104} (\bibinfo {year} {2019})}\BibitemShut
  {NoStop}%
\bibitem [{\citenamefont {Xu}\ \emph {et~al.}(2007)\citenamefont {Xu},
  \citenamefont {Dong},\ and\ \citenamefont {Lipson}}]{Xu2007d}%
  \BibitemOpen
  \bibfield  {author} {\bibinfo {author} {\bibfnamefont {Q.}~\bibnamefont
  {Xu}}, \bibinfo {author} {\bibfnamefont {P.}~\bibnamefont {Dong}}, \ and\
  \bibinfo {author} {\bibfnamefont {M.}~\bibnamefont {Lipson}},\ }\href
  {http://www.nature.com/doifinder/10.1038/nphys600} {\bibfield  {journal}
  {\bibinfo  {journal} {Nat. Phys.}\ }\textbf {\bibinfo {volume} {3}},\
  \bibinfo {pages} {406} (\bibinfo {year} {2007})}\BibitemShut {NoStop}%
\bibitem [{\citenamefont {Preble}\ \emph {et~al.}(2007)\citenamefont {Preble},
  \citenamefont {Xu},\ and\ \citenamefont {Lipson}}]{Preble2007}%
  \BibitemOpen
  \bibfield  {author} {\bibinfo {author} {\bibfnamefont {S.~F.}\ \bibnamefont
  {Preble}}, \bibinfo {author} {\bibfnamefont {Q.}~\bibnamefont {Xu}}, \ and\
  \bibinfo {author} {\bibfnamefont {M.}~\bibnamefont {Lipson}},\ }\href@noop {}
  {\bibfield  {journal} {\bibinfo  {journal} {Nature Photon.}\ }\textbf
  {\bibinfo {volume} {1}},\ \bibinfo {pages} {293} (\bibinfo {year}
  {2007})}\BibitemShut {NoStop}%
\bibitem [{\citenamefont {Tanabe}\ \emph {et~al.}(2009)\citenamefont {Tanabe},
  \citenamefont {Notomi}, \citenamefont {Taniyama},\ and\ \citenamefont
  {Kuramochi}}]{Tanabe2009}%
  \BibitemOpen
  \bibfield  {author} {\bibinfo {author} {\bibfnamefont {T.}~\bibnamefont
  {Tanabe}}, \bibinfo {author} {\bibfnamefont {M.}~\bibnamefont {Notomi}},
  \bibinfo {author} {\bibfnamefont {H.}~\bibnamefont {Taniyama}}, \ and\
  \bibinfo {author} {\bibfnamefont {E.}~\bibnamefont {Kuramochi}},\ }\href
  {\doibase 10.1103/PhysRevLett.102.043907} {\bibfield  {journal} {\bibinfo
  {journal} {Phys. Rev. Lett.}\ }\textbf {\bibinfo {volume} {102}},\ \bibinfo
  {pages} {043907} (\bibinfo {year} {2009})}\BibitemShut {NoStop}%
\bibitem [{\citenamefont {Lee}\ \emph {et~al.}(2018)\citenamefont {Lee},
  \citenamefont {Son}, \citenamefont {Kang}, \citenamefont {Park},
  \citenamefont {Rotermund},\ and\ \citenamefont {Min}}]{Lee2018}%
  \BibitemOpen
  \bibfield  {author} {\bibinfo {author} {\bibfnamefont {K.}~\bibnamefont
  {Lee}}, \bibinfo {author} {\bibfnamefont {J.}~\bibnamefont {Son}}, \bibinfo
  {author} {\bibfnamefont {B.}~\bibnamefont {Kang}}, \bibinfo {author}
  {\bibfnamefont {J.}~\bibnamefont {Park}}, \bibinfo {author} {\bibfnamefont
  {F.}~\bibnamefont {Rotermund}}, \ and\ \bibinfo {author} {\bibfnamefont
  {B.}~\bibnamefont {Min}},\ }\href {\doibase 10.1109/IRMMW-THz.2017.8067111}
  {\bibfield  {journal} {\bibinfo  {journal} {Nature Photon.}\ }\textbf
  {\bibinfo {volume} {12}},\ \bibinfo {pages} {765} (\bibinfo {year}
  {2018})}\BibitemShut {NoStop}%
\bibitem [{\citenamefont {Shcherbakov}\ \emph {et~al.}(2019)\citenamefont
  {Shcherbakov}, \citenamefont {Werner}, \citenamefont {Fan}, \citenamefont
  {Talisa}, \citenamefont {Chowdhury},\ and\ \citenamefont
  {Shvets}}]{Shcherbakov2019}%
  \BibitemOpen
  \bibfield  {author} {\bibinfo {author} {\bibfnamefont {M.~R.}\ \bibnamefont
  {Shcherbakov}}, \bibinfo {author} {\bibfnamefont {K.}~\bibnamefont {Werner}},
  \bibinfo {author} {\bibfnamefont {Z.}~\bibnamefont {Fan}}, \bibinfo {author}
  {\bibfnamefont {N.}~\bibnamefont {Talisa}}, \bibinfo {author} {\bibfnamefont
  {E.}~\bibnamefont {Chowdhury}}, \ and\ \bibinfo {author} {\bibfnamefont
  {G.}~\bibnamefont {Shvets}},\ }\href {\doibase 10.1038/s41467-019-09313-8}
 {\bibfield  {journal} {\bibinfo  {journal}
 		{Nat. Commun.}\ }\textbf {\bibinfo {volume} {10}},\ \bibinfo {pages}
 	{1345} (\bibinfo {year} {2019})}\BibitemShut {NoStop}%
\bibitem [{\citenamefont {Shaltout}\ \emph {et~al.}(2015)\citenamefont
  {Shaltout}, \citenamefont {Kildishev},\ and\ \citenamefont
  {Shalaev}}]{Shaltout2015}%
  \BibitemOpen
  \bibfield  {author} {\bibinfo {author} {\bibfnamefont {A.}~\bibnamefont
  {Shaltout}}, \bibinfo {author} {\bibfnamefont {A.}~\bibnamefont {Kildishev}},
  \ and\ \bibinfo {author} {\bibfnamefont {V.}~\bibnamefont {Shalaev}},\ }\href
  {\doibase 10.1364/OME.5.002459} {\bibfield  {journal} {\bibinfo  {journal}
  {Opt. Mater. Express}\ }\textbf {\bibinfo {volume} {5}},\ \bibinfo {pages}
  {2459} (\bibinfo {year} {2015})}\BibitemShut {NoStop}%
\bibitem [{\citenamefont {Sounas}\ and\ \citenamefont
  {Al{\`{u}}}(2017)}]{Sounas2017}%
  \BibitemOpen
  \bibfield  {author} {\bibinfo {author} {\bibfnamefont {D.~L.}\ \bibnamefont
  {Sounas}}\ and\ \bibinfo {author} {\bibfnamefont {A.}~\bibnamefont
  {Al{\`{u}}}},\ }\href {\doibase 10.1038/s41566-017-0051-x} {\bibfield
  {journal} {\bibinfo  {journal} {Nature Photon.}\ }\textbf {\bibinfo {volume}
  {11}},\ \bibinfo {pages} {774} (\bibinfo {year} {2017})}\BibitemShut
  {NoStop}%
\bibitem [{\citenamefont {Caloz}\ \emph {et~al.}(2018)\citenamefont {Caloz},
  \citenamefont {Al{\`{u}}}, \citenamefont {Tretyakov}, \citenamefont
  {Sounas},\ and\ \citenamefont {Achouri}}]{Caloz2018a}%
  \BibitemOpen
  \bibfield  {author} {\bibinfo {author} {\bibfnamefont {C.}~\bibnamefont
  {Caloz}}, \bibinfo {author} {\bibfnamefont {A.}~\bibnamefont {Al{\`{u}}}},
  \bibinfo {author} {\bibfnamefont {S.}~\bibnamefont {Tretyakov}}, \bibinfo
  {author} {\bibfnamefont {D.}~\bibnamefont {Sounas}}, \ and\ \bibinfo {author}
  {\bibfnamefont {K.}~\bibnamefont {Achouri}},\ }\href {\doibase
  10.1103/PhysRevApplied.10.047001} {\bibfield  {journal} {\bibinfo  {journal}
  {Physical Review Applied}\ }\textbf {\bibinfo {volume} {10}},\ \bibinfo
  {pages} {047001} (\bibinfo {year} {2018})}\BibitemShut {NoStop}%
\bibitem [{\citenamefont {Minkov}\ and\ \citenamefont
  {Savona}(2016)}]{Minkov2016}%
  \BibitemOpen
  \bibfield  {author} {\bibinfo {author} {\bibfnamefont {M.}~\bibnamefont
  {Minkov}}\ and\ \bibinfo {author} {\bibfnamefont {V.}~\bibnamefont
  {Savona}},\ }\href {\doibase 10.1364/OPTICA.3.000200} {\bibfield  {journal}
  {\bibinfo  {journal} {Optica}\ }\textbf {\bibinfo {volume} {3}},\ \bibinfo
  {pages} {200} (\bibinfo {year} {2016})}\BibitemShut {NoStop}%
\bibitem [{\citenamefont {Rogov}\ and\ \citenamefont
  {Narimanov}(2018)}]{Rogov2018}%
  \BibitemOpen
  \bibfield  {author} {\bibinfo {author} {\bibfnamefont {A.}~\bibnamefont
  {Rogov}}\ and\ \bibinfo {author} {\bibfnamefont {E.}~\bibnamefont
  {Narimanov}},\ }\href {\doibase 10.1021/acsphotonics.8b00233} {\bibfield
  {journal} {\bibinfo  {journal} {ACS Photonics}\ }\textbf {\bibinfo {volume}
  {5}},\ \bibinfo {pages} {2868} (\bibinfo {year} {2018})}\BibitemShut
  {NoStop}%
\bibitem [{\citenamefont {Qu}\ \emph {et~al.}(2018)\citenamefont {Qu},
  \citenamefont {Jia}, \citenamefont {Edwards},\ and\ \citenamefont
  {Fisch}}]{Qu2018a}%
  \BibitemOpen
  \bibfield  {author} {\bibinfo {author} {\bibfnamefont {K.}~\bibnamefont
  {Qu}}, \bibinfo {author} {\bibfnamefont {Q.}~\bibnamefont {Jia}}, \bibinfo
  {author} {\bibfnamefont {M.~R.}\ \bibnamefont {Edwards}}, \ and\ \bibinfo
  {author} {\bibfnamefont {N.~J.}\ \bibnamefont {Fisch}},\ }\href {\doibase
  10.1103/PhysRevE.98.023202} {\bibfield  {journal} {\bibinfo  {journal} {Phys.
  Rev. E}\ }\textbf {\bibinfo {volume} {98}},\ \bibinfo {pages} {023202}
  (\bibinfo {year} {2018})}\BibitemShut {NoStop}%
\bibitem [{\citenamefont {Haus}(1984)}]{Haus1984}%
  \BibitemOpen
  \bibfield  {author} {\bibinfo {author} {\bibfnamefont {H.~A.}\ \bibnamefont
  {Haus}},\ }\href@noop {} {\emph {\bibinfo {title} {{Waves And Fields In
  Optoelectronics}}}}\ (\bibinfo  {publisher} {Prentice-Hall},\ \bibinfo
  {address} {New Jersey},\ \bibinfo {year} {1984})\BibitemShut {NoStop}%
\bibitem [{\citenamefont {Fan}\ \emph {et~al.}(2003)\citenamefont {Fan},
  \citenamefont {Suh},\ and\ \citenamefont {Joannopoulos}}]{Fan2003a}%
  \BibitemOpen
  \bibfield  {author} {\bibinfo {author} {\bibfnamefont {S.~F.}\ \bibnamefont
  {Fan}}, \bibinfo {author} {\bibfnamefont {W.}~\bibnamefont {Suh}}, \ and\
  \bibinfo {author} {\bibfnamefont {J.~D.}\ \bibnamefont {Joannopoulos}},\
  }\href {\doibase 10.1364/JOSAA.20.000569} {\bibfield  {journal} {\bibinfo
  {journal} {J. Opt. Soc. Am. A}\ }\textbf {\bibinfo {volume} {20}},\ \bibinfo
  {pages} {569} (\bibinfo {year} {2003})}\BibitemShut {NoStop}%
\bibitem [{\citenamefont {Minkov}\ \emph {et~al.}(2017)\citenamefont {Minkov},
  \citenamefont {Shi},\ and\ \citenamefont {Fan}}]{Minkov2017}%
  \BibitemOpen
  \bibfield  {author} {\bibinfo {author} {\bibfnamefont {M.}~\bibnamefont
  {Minkov}}, \bibinfo {author} {\bibfnamefont {Y.}~\bibnamefont {Shi}}, \ and\
  \bibinfo {author} {\bibfnamefont {S.}~\bibnamefont {Fan}},\ }\href {\doibase
  10.1063/1.4985381} {\bibfield  {journal} {\bibinfo  {journal} {APL
  Photonics}\ }\textbf {\bibinfo {volume} {2}},\ \bibinfo {pages} {076101}
  (\bibinfo {year} {2017})}\BibitemShut {NoStop}%
\bibitem [{\citenamefont {Boyd}(2008)}]{Boyd2008}%
  \BibitemOpen
  \bibfield  {author} {\bibinfo {author} {\bibfnamefont {R.~W.}\ \bibnamefont
  {Boyd}},\ }\href@noop {} {\emph {\bibinfo {title} {{Nonlinear Optics}}}},\
  \bibinfo {edition} {3rd}\ ed.\ (\bibinfo  {publisher} {Academic Press},\
  \bibinfo {year} {2008})\BibitemShut {NoStop}%
\bibitem [{\citenamefont {Lu}\ \emph {et~al.}(2014)\citenamefont {Lu},
  \citenamefont {Joannopoulos},\ and\ \citenamefont {Soljacic}}]{Lu2014}%
  \BibitemOpen
  \bibfield  {author} {\bibinfo {author} {\bibfnamefont {L.}~\bibnamefont
  {Lu}}, \bibinfo {author} {\bibfnamefont {J.~D.}\ \bibnamefont
  {Joannopoulos}}, \ and\ \bibinfo {author} {\bibfnamefont {M.}~\bibnamefont
  {Soljacic}},\ }\href {\doibase 10.1038/nphoton.2014.248} {\bibfield
  {journal} {\bibinfo  {journal} {Nature Photon.}\ }\textbf {\bibinfo {volume}
  {8}},\ \bibinfo {pages} {821} (\bibinfo {year} {2014})}\BibitemShut {NoStop}%
\bibitem [{\citenamefont {Khanikaev}\ and\ \citenamefont
  {Shvets}(2017)}]{Khanikaev2017}%
  \BibitemOpen
  \bibfield  {author} {\bibinfo {author} {\bibfnamefont {A.~B.}\ \bibnamefont
  {Khanikaev}}\ and\ \bibinfo {author} {\bibfnamefont {G.}~\bibnamefont
  {Shvets}},\ }\href {\doibase 10.1038/s41566-017-0048-5} {\bibfield  {journal}
  {\bibinfo  {journal} {Nature Photon.}\ }\textbf {\bibinfo {volume} {11}},\
  \bibinfo {pages} {763} (\bibinfo {year} {2017})}\BibitemShut {NoStop}%
\bibitem [{\citenamefont {Strickland}\ and\ \citenamefont
  {Mourou}(1985)}]{Strickland1985}%
  \BibitemOpen
  \bibfield  {author} {\bibinfo {author} {\bibfnamefont {D.}~\bibnamefont
  {Strickland}}\ and\ \bibinfo {author} {\bibfnamefont {G.}~\bibnamefont
  {Mourou}},\ }\href {\doibase 10.1016/0030-4018(85)90120-8} {\bibfield
  {journal} {\bibinfo  {journal} {Opt. Commun.}\ }\textbf {\bibinfo {volume}
  {56}},\ \bibinfo {pages} {219} (\bibinfo {year} {1985})}\BibitemShut
  {NoStop}%
\bibitem [{\citenamefont {Weiner}(2011)}]{Weiner2011}%
  \BibitemOpen
  \bibfield  {author} {\bibinfo {author} {\bibfnamefont {A.~M.}\ \bibnamefont
  {Weiner}},\ }\href {\doibase 10.1016/j.optcom.2011.03.084} {\bibfield
  {journal} {\bibinfo  {journal} {Opt. Commun.}\ }\textbf {\bibinfo {volume}
  {284}},\ \bibinfo {pages} {3669} (\bibinfo {year} {2011})}\BibitemShut
  {NoStop}%
\bibitem [{\citenamefont {Wilks}\ \emph {et~al.}(1989)\citenamefont {Wilks},
  \citenamefont {Dawson}, \citenamefont {Mori}, \citenamefont {Katsouleas},\
  and\ \citenamefont {Jones}}]{Wilks1989}%
  \BibitemOpen
  \bibfield  {author} {\bibinfo {author} {\bibfnamefont {S.~C.}\ \bibnamefont
  {Wilks}}, \bibinfo {author} {\bibfnamefont {J.~M.}\ \bibnamefont {Dawson}},
  \bibinfo {author} {\bibfnamefont {W.~B.}\ \bibnamefont {Mori}}, \bibinfo
  {author} {\bibfnamefont {T.}~\bibnamefont {Katsouleas}}, \ and\ \bibinfo
  {author} {\bibfnamefont {M.~E.}\ \bibnamefont {Jones}},\ }\href {\doibase
  10.1103/PhysRevLett.62.2600} {\bibfield  {journal} {\bibinfo  {journal}
  {Phys. Rev. Lett.}\ }\textbf {\bibinfo {volume} {62}},\ \bibinfo {pages}
  {2600} (\bibinfo {year} {1989})}\BibitemShut {NoStop}%
\bibitem [{\citenamefont {Siders}\ \emph {et~al.}(1996)\citenamefont {Siders},
  \citenamefont {Turner}, \citenamefont {Downer}, \citenamefont {Babine},
  \citenamefont {Stepanov},\ and\ \citenamefont {Sergeev}}]{Siders1996}%
  \BibitemOpen
  \bibfield  {author} {\bibinfo {author} {\bibfnamefont {C.~W.}\ \bibnamefont
  {Siders}}, \bibinfo {author} {\bibfnamefont {N.~C.}\ \bibnamefont {Turner}},
  \bibinfo {author} {\bibfnamefont {M.~C.}\ \bibnamefont {Downer}}, \bibinfo
  {author} {\bibfnamefont {A.}~\bibnamefont {Babine}}, \bibinfo {author}
  {\bibfnamefont {A.}~\bibnamefont {Stepanov}}, \ and\ \bibinfo {author}
  {\bibfnamefont {A.~M.}\ \bibnamefont {Sergeev}},\ }\href {\doibase
  10.1364/josab.13.000330} {\bibfield  {journal} {\bibinfo  {journal} {J. Opt.
  Soc. Am. B}\ }\textbf {\bibinfo {volume} {13}},\ \bibinfo {pages} {330}
  (\bibinfo {year} {1996})}\BibitemShut {NoStop}%
\bibitem [{\citenamefont {Han}\ \emph {et~al.}(2016)\citenamefont {Han},
  \citenamefont {Kim}, \citenamefont {Kim}, \citenamefont {Kim}, \citenamefont
  {Kim}, \citenamefont {Park},\ and\ \citenamefont {Kim}}]{Han2016}%
  \BibitemOpen
  \bibfield  {author} {\bibinfo {author} {\bibfnamefont {S.}~\bibnamefont
  {Han}}, \bibinfo {author} {\bibfnamefont {H.}~\bibnamefont {Kim}}, \bibinfo
  {author} {\bibfnamefont {Y.~W.}\ \bibnamefont {Kim}}, \bibinfo {author}
  {\bibfnamefont {Y.-J.}\ \bibnamefont {Kim}}, \bibinfo {author} {\bibfnamefont
  {S.}~\bibnamefont {Kim}}, \bibinfo {author} {\bibfnamefont {I.-Y.}\
  \bibnamefont {Park}}, \ and\ \bibinfo {author} {\bibfnamefont {S.-W.}\
  \bibnamefont {Kim}},\ }\href {\doibase 10.1038/ncomms13105} {\bibfield
  {journal} {\bibinfo  {journal} {Nat. Commun.}\ }\textbf {\bibinfo {volume}
  {7}},\ \bibinfo {pages} {13105} (\bibinfo {year} {2016})}\BibitemShut
  {NoStop}%
\bibitem [{\citenamefont {Liu}\ \emph {et~al.}(2018{\natexlab{a}})\citenamefont
  {Liu}, \citenamefont {Guo}, \citenamefont {Vampa}, \citenamefont {Zhang},
  \citenamefont {Sarmiento}, \citenamefont {Xiao}, \citenamefont {Bucksbaum},
  \citenamefont {Vu{\v{c}}kovi{\'{c}}}, \citenamefont {Fan},\ and\
  \citenamefont {Reis}}]{Liu2018f}%
  \BibitemOpen
  \bibfield  {author} {\bibinfo {author} {\bibfnamefont {H.}~\bibnamefont
  {Liu}}, \bibinfo {author} {\bibfnamefont {C.}~\bibnamefont {Guo}}, \bibinfo
  {author} {\bibfnamefont {G.}~\bibnamefont {Vampa}}, \bibinfo {author}
  {\bibfnamefont {J.~L.}\ \bibnamefont {Zhang}}, \bibinfo {author}
  {\bibfnamefont {T.}~\bibnamefont {Sarmiento}}, \bibinfo {author}
  {\bibfnamefont {M.}~\bibnamefont {Xiao}}, \bibinfo {author} {\bibfnamefont
  {P.~H.}\ \bibnamefont {Bucksbaum}}, \bibinfo {author} {\bibfnamefont
  {J.}~\bibnamefont {Vu{\v{c}}kovi{\'{c}}}}, \bibinfo {author} {\bibfnamefont
  {S.}~\bibnamefont {Fan}}, \ and\ \bibinfo {author} {\bibfnamefont {D.~A.}\
  \bibnamefont {Reis}},\ }\href {\doibase 10.1038/s41567-018-0233-6} {\bibfield
   {journal} {\bibinfo  {journal} {Nat. Phys.}\ }\textbf {\bibinfo {volume}
  {14}},\ \bibinfo {pages} {1006} (\bibinfo {year}
  {2018}{\natexlab{a}})}\BibitemShut {NoStop}%
\bibitem [{\citenamefont {Ghimire}\ and\ \citenamefont
  {Reis}(2019)}]{Ghimire2019}%
  \BibitemOpen
  \bibfield  {author} {\bibinfo {author} {\bibfnamefont {S.}~\bibnamefont
  {Ghimire}}\ and\ \bibinfo {author} {\bibfnamefont {D.~A.}\ \bibnamefont
  {Reis}},\ }\href {\doibase 10.1038/s41567-018-0315-5} {\bibfield  {journal}
  {\bibinfo  {journal} {Nat. Phys.}\ }\textbf {\bibinfo {volume} {15}},\
  \bibinfo {pages} {10} (\bibinfo {year} {2019})}\BibitemShut {NoStop}%
\bibitem [{\citenamefont {Lanin}\ \emph {et~al.}(2017)\citenamefont {Lanin},
  \citenamefont {Stepanov}, \citenamefont {Fedotov},\ and\ \citenamefont
  {Zheltikov}}]{Anin2017}%
  \BibitemOpen
  \bibfield  {author} {\bibinfo {author} {\bibfnamefont {A.~A.}\ \bibnamefont
  {Lanin}}, \bibinfo {author} {\bibfnamefont {E.~A.}\ \bibnamefont {Stepanov}},
  \bibinfo {author} {\bibfnamefont {A.~B.}\ \bibnamefont {Fedotov}}, \ and\
  \bibinfo {author} {\bibfnamefont {A.~M.}\ \bibnamefont {Zheltikov}},\
  }\href@noop {} {\bibfield  {journal} {\bibinfo  {journal} {Optica}\ }\textbf
  {\bibinfo {volume} {4}},\ \bibinfo {pages} {516} (\bibinfo {year}
  {2017})}\BibitemShut {NoStop}%
\bibitem [{\citenamefont {Gariepy}\ \emph {et~al.}(2014)\citenamefont
  {Gariepy}, \citenamefont {Leach}, \citenamefont {Kim}, \citenamefont
  {Hammond}, \citenamefont {Frumker}, \citenamefont {Boyd},\ and\ \citenamefont
  {Corkum}}]{Gariepy2014}%
  \BibitemOpen
  \bibfield  {author} {\bibinfo {author} {\bibfnamefont {G.}~\bibnamefont
  {Gariepy}}, \bibinfo {author} {\bibfnamefont {J.}~\bibnamefont {Leach}},
  \bibinfo {author} {\bibfnamefont {K.~T.}\ \bibnamefont {Kim}}, \bibinfo
  {author} {\bibfnamefont {T.~J.}\ \bibnamefont {Hammond}}, \bibinfo {author}
  {\bibfnamefont {E.}~\bibnamefont {Frumker}}, \bibinfo {author} {\bibfnamefont
  {R.~W.}\ \bibnamefont {Boyd}}, \ and\ \bibinfo {author} {\bibfnamefont
  {P.~B.}\ \bibnamefont {Corkum}},\ }\href {\doibase
  10.1103/PhysRevLett.113.153901} {\bibfield  {journal} {\bibinfo  {journal}
  {Phys. Rev. Lett.}\ }\textbf {\bibinfo {volume} {113}},\ \bibinfo {pages}
  {153901} (\bibinfo {year} {2014})}\BibitemShut {NoStop}%
\bibitem [{\citenamefont {Li}\ \emph {et~al.}(2017)\citenamefont {Li},
  \citenamefont {Brown}, \citenamefont {Ko}, \citenamefont {Kong},\ and\
  \citenamefont {Zhang}}]{Li2017c}%
  \BibitemOpen
  \bibfield  {author} {\bibinfo {author} {\bibfnamefont {Z.}~\bibnamefont
  {Li}}, \bibinfo {author} {\bibfnamefont {G.~G.}\ \bibnamefont {Brown}},
  \bibinfo {author} {\bibfnamefont {D.~H.}\ \bibnamefont {Ko}}, \bibinfo
  {author} {\bibfnamefont {F.}~\bibnamefont {Kong}}, \ and\ \bibinfo {author}
  {\bibfnamefont {C.}~\bibnamefont {Zhang}},\ }\href {\doibase
  10.1038/ncomms14970} {\bibfield  {journal} {\bibinfo  {journal} {Nat.
  Commun.}\ }\textbf {\bibinfo {volume} {8}},\ \bibinfo {pages} {14970}
  (\bibinfo {year} {2017})}\BibitemShut {NoStop}%
\bibitem [{\citenamefont {Wagner}\ and\ \citenamefont
  {Harned}(2010)}]{Wagner2010}%
  \BibitemOpen
  \bibfield  {author} {\bibinfo {author} {\bibfnamefont {C.}~\bibnamefont
  {Wagner}}\ and\ \bibinfo {author} {\bibfnamefont {N.}~\bibnamefont
  {Harned}},\ }\href {\doibase 10.1038/nphoton.2009.251} {\bibfield  {journal}
  {\bibinfo  {journal} {Nature Photon.}\ }\textbf {\bibinfo {volume} {4}},\
  \bibinfo {pages} {24} (\bibinfo {year} {2010})}\BibitemShut {NoStop}%
\bibitem [{\citenamefont {Seaberg}\ \emph {et~al.}(2011)\citenamefont
  {Seaberg}, \citenamefont {Adams}, \citenamefont {Townsend}, \citenamefont
  {Raymondson}, \citenamefont {Schlotter}, \citenamefont {Liu}, \citenamefont
  {Menoni}, \citenamefont {Rong}, \citenamefont {Chen}, \citenamefont {Miao},
  \citenamefont {Kapteyn},\ and\ \citenamefont {Murnane}}]{Seaberg2011}%
  \BibitemOpen
  \bibfield  {author} {\bibinfo {author} {\bibfnamefont {M.~D.}\ \bibnamefont
  {Seaberg}}, \bibinfo {author} {\bibfnamefont {D.~E.}\ \bibnamefont {Adams}},
  \bibinfo {author} {\bibfnamefont {E.~L.}\ \bibnamefont {Townsend}}, \bibinfo
  {author} {\bibfnamefont {D.~A.}\ \bibnamefont {Raymondson}}, \bibinfo
  {author} {\bibfnamefont {W.~F.}\ \bibnamefont {Schlotter}}, \bibinfo {author}
  {\bibfnamefont {Y.}~\bibnamefont {Liu}}, \bibinfo {author} {\bibfnamefont
  {C.~S.}\ \bibnamefont {Menoni}}, \bibinfo {author} {\bibfnamefont
  {L.}~\bibnamefont {Rong}}, \bibinfo {author} {\bibfnamefont {C.-C.}\
  \bibnamefont {Chen}}, \bibinfo {author} {\bibfnamefont {J.}~\bibnamefont
  {Miao}}, \bibinfo {author} {\bibfnamefont {H.~C.}\ \bibnamefont {Kapteyn}}, \
  and\ \bibinfo {author} {\bibfnamefont {M.~M.}\ \bibnamefont {Murnane}},\
  }\href@noop {} {\bibfield  {journal} {\bibinfo  {journal} {Opt. Express}\
  }\textbf {\bibinfo {volume} {19}},\ \bibinfo {pages} {7235} (\bibinfo {year}
  {2011})}\BibitemShut {NoStop}%
\bibitem [{\citenamefont {Ghimire}\ \emph {et~al.}(2011)\citenamefont
  {Ghimire}, \citenamefont {Dichiara}, \citenamefont {Sistrunk}, \citenamefont
  {Agostini}, \citenamefont {Dimauro},\ and\ \citenamefont
  {Reis}}]{Ghimire2011}%
  \BibitemOpen
  \bibfield  {author} {\bibinfo {author} {\bibfnamefont {S.}~\bibnamefont
  {Ghimire}}, \bibinfo {author} {\bibfnamefont {A.~D.}\ \bibnamefont
  {Dichiara}}, \bibinfo {author} {\bibfnamefont {E.}~\bibnamefont {Sistrunk}},
  \bibinfo {author} {\bibfnamefont {P.}~\bibnamefont {Agostini}}, \bibinfo
  {author} {\bibfnamefont {L.~F.}\ \bibnamefont {Dimauro}}, \ and\ \bibinfo
  {author} {\bibfnamefont {D.~A.}\ \bibnamefont {Reis}},\ }\href {\doibase
  10.1038/nphys1847} {\bibfield  {journal} {\bibinfo  {journal} {Nat. Phys.}\
  }\textbf {\bibinfo {volume} {7}},\ \bibinfo {pages} {138} (\bibinfo {year}
  {2011})}\BibitemShut {NoStop}%
\bibitem [{\citenamefont {Vampa}\ \emph {et~al.}(2014)\citenamefont {Vampa},
  \citenamefont {McDonald}, \citenamefont {Orlando}, \citenamefont {Klug},
  \citenamefont {Corkum},\ and\ \citenamefont {Brabec}}]{Vampa2014a}%
  \BibitemOpen
  \bibfield  {author} {\bibinfo {author} {\bibfnamefont {G.}~\bibnamefont
  {Vampa}}, \bibinfo {author} {\bibfnamefont {C.~R.}\ \bibnamefont {McDonald}},
  \bibinfo {author} {\bibfnamefont {G.}~\bibnamefont {Orlando}}, \bibinfo
  {author} {\bibfnamefont {D.~D.}\ \bibnamefont {Klug}}, \bibinfo {author}
  {\bibfnamefont {P.~B.}\ \bibnamefont {Corkum}}, \ and\ \bibinfo {author}
  {\bibfnamefont {T.}~\bibnamefont {Brabec}},\ }\href {\doibase
  10.1103/PhysRevLett.113.073901} {\bibfield  {journal} {\bibinfo  {journal}
  {Phys. Rev. Lett.}\ }\textbf {\bibinfo {volume} {113}},\ \bibinfo {pages}
  {073901} (\bibinfo {year} {2014})}\BibitemShut {NoStop}%
\bibitem [{\citenamefont {Vampa}\ \emph {et~al.}(2015)\citenamefont {Vampa},
  \citenamefont {Mcdonald}, \citenamefont {Orlando}, \citenamefont {Corkum},\
  and\ \citenamefont {Brabec}}]{Vampa2015}%
  \BibitemOpen
  \bibfield  {author} {\bibinfo {author} {\bibfnamefont {G.}~\bibnamefont
  {Vampa}}, \bibinfo {author} {\bibfnamefont {C.~R.}\ \bibnamefont {Mcdonald}},
  \bibinfo {author} {\bibfnamefont {G.}~\bibnamefont {Orlando}}, \bibinfo
  {author} {\bibfnamefont {P.~B.}\ \bibnamefont {Corkum}}, \ and\ \bibinfo
  {author} {\bibfnamefont {T.}~\bibnamefont {Brabec}},\ }\href {\doibase
  10.1103/PhysRevB.91.064302} {\bibfield  {journal} {\bibinfo  {journal} {Phys.
  Rev. B}\ }\textbf {\bibinfo {volume} {91}},\ \bibinfo {pages} {064302}
  (\bibinfo {year} {2015})}\BibitemShut {NoStop}%
\bibitem [{\citenamefont {Wolter}\ \emph {et~al.}(2015)\citenamefont {Wolter},
  \citenamefont {Pullen}, \citenamefont {Baudisch}, \citenamefont {Sclafani},
  \citenamefont {Hemmer}, \citenamefont {Senftleben}, \citenamefont
  {Schr{\"{o}}ter}, \citenamefont {Ullrich}, \citenamefont {Moshammer},\ and\
  \citenamefont {Biegert}}]{Wolter2015}%
  \BibitemOpen
  \bibfield  {author} {\bibinfo {author} {\bibfnamefont {B.}~\bibnamefont
  {Wolter}}, \bibinfo {author} {\bibfnamefont {M.~G.}\ \bibnamefont {Pullen}},
  \bibinfo {author} {\bibfnamefont {M.}~\bibnamefont {Baudisch}}, \bibinfo
  {author} {\bibfnamefont {M.}~\bibnamefont {Sclafani}}, \bibinfo {author}
  {\bibfnamefont {M.}~\bibnamefont {Hemmer}}, \bibinfo {author} {\bibfnamefont
  {A.}~\bibnamefont {Senftleben}}, \bibinfo {author} {\bibfnamefont {C.~D.}\
  \bibnamefont {Schr{\"{o}}ter}}, \bibinfo {author} {\bibfnamefont
  {J.}~\bibnamefont {Ullrich}}, \bibinfo {author} {\bibfnamefont
  {R.}~\bibnamefont {Moshammer}}, \ and\ \bibinfo {author} {\bibfnamefont
  {J.}~\bibnamefont {Biegert}},\ }\href {\doibase 10.1103/PhysRevX.5.021034}
  {\bibfield  {journal} {\bibinfo  {journal} {Phys. Rev. X}\ }\textbf {\bibinfo
  {volume} {5}},\ \bibinfo {pages} {021034} (\bibinfo {year}
  {2015})}\BibitemShut {NoStop}%
\bibitem [{\citenamefont {You}\ \emph {et~al.}(2017)\citenamefont {You},
  \citenamefont {Yin}, \citenamefont {Wu}, \citenamefont {Chew}, \citenamefont
  {Ren}, \citenamefont {Zhuang}, \citenamefont {Gholam-Mirzaei}, \citenamefont
  {Chini}, \citenamefont {Chang},\ and\ \citenamefont {Ghimire}}]{You2017}%
  \BibitemOpen
  \bibfield  {author} {\bibinfo {author} {\bibfnamefont {Y.~S.}\ \bibnamefont
  {You}}, \bibinfo {author} {\bibfnamefont {Y.}~\bibnamefont {Yin}}, \bibinfo
  {author} {\bibfnamefont {Y.}~\bibnamefont {Wu}}, \bibinfo {author}
  {\bibfnamefont {A.}~\bibnamefont {Chew}}, \bibinfo {author} {\bibfnamefont
  {X.}~\bibnamefont {Ren}}, \bibinfo {author} {\bibfnamefont {F.}~\bibnamefont
  {Zhuang}}, \bibinfo {author} {\bibfnamefont {S.}~\bibnamefont
  {Gholam-Mirzaei}}, \bibinfo {author} {\bibfnamefont {M.}~\bibnamefont
  {Chini}}, \bibinfo {author} {\bibfnamefont {Z.}~\bibnamefont {Chang}}, \ and\
  \bibinfo {author} {\bibfnamefont {S.}~\bibnamefont {Ghimire}},\ }\href
  {\doibase 10.1038/s41467-017-00989-4} {\bibfield  {journal} {\bibinfo
  {journal} {Nat. Commun.}\ }\textbf {\bibinfo {volume} {8}},\ \bibinfo {pages}
  {724} (\bibinfo {year} {2017})}\BibitemShut {NoStop}%
\bibitem [{\citenamefont {Preble}\ \emph {et~al.}(2012)\citenamefont {Preble},
  \citenamefont {Cao}, \citenamefont {Elshaari}, \citenamefont {Aboketaf},\
  and\ \citenamefont {Adams}}]{Preble2012}%
  \BibitemOpen
  \bibfield  {author} {\bibinfo {author} {\bibfnamefont {S.}~\bibnamefont
  {Preble}}, \bibinfo {author} {\bibfnamefont {L.}~\bibnamefont {Cao}},
  \bibinfo {author} {\bibfnamefont {A.}~\bibnamefont {Elshaari}}, \bibinfo
  {author} {\bibfnamefont {A.}~\bibnamefont {Aboketaf}}, \ and\ \bibinfo
  {author} {\bibfnamefont {D.}~\bibnamefont {Adams}},\ }\href {\doibase
  10.1063/1.4764068} {\bibfield  {journal} {\bibinfo  {journal} {Appl. Phys.
  Lett.}\ }\textbf {\bibinfo {volume} {101}},\ \bibinfo {pages} {171110}
  (\bibinfo {year} {2012})}\BibitemShut {NoStop}%
\bibitem [{\citenamefont {Notomi}\ \emph {et~al.}(2007)\citenamefont {Notomi},
  \citenamefont {Tanabe}, \citenamefont {Shinya}, \citenamefont {Kuramochi},
  \citenamefont {Taniyama}, \citenamefont {Mitsugi},\ and\ \citenamefont
  {Morita}}]{Notomi2007a}%
  \BibitemOpen
  \bibfield  {author} {\bibinfo {author} {\bibfnamefont {M.}~\bibnamefont
  {Notomi}}, \bibinfo {author} {\bibfnamefont {T.}~\bibnamefont {Tanabe}},
  \bibinfo {author} {\bibfnamefont {A.}~\bibnamefont {Shinya}}, \bibinfo
  {author} {\bibfnamefont {E.}~\bibnamefont {Kuramochi}}, \bibinfo {author}
  {\bibfnamefont {H.}~\bibnamefont {Taniyama}}, \bibinfo {author}
  {\bibfnamefont {S.}~\bibnamefont {Mitsugi}}, \ and\ \bibinfo {author}
  {\bibfnamefont {M.}~\bibnamefont {Morita}},\ }\href {\doibase
  10.1364/OE.15.017458} {\bibfield  {journal} {\bibinfo  {journal} {Opt.
  Express}\ }\textbf {\bibinfo {volume} {15}},\ \bibinfo {pages} {17458}
  (\bibinfo {year} {2007})}\BibitemShut {NoStop}%
\bibitem [{\citenamefont {Bennett}\ \emph {et~al.}(1990)\citenamefont
  {Bennett}, \citenamefont {Soref},\ and\ \citenamefont {{Del
  Alamo}}}]{Bennett1990}%
  \BibitemOpen
  \bibfield  {author} {\bibinfo {author} {\bibfnamefont {B.~R.}\ \bibnamefont
  {Bennett}}, \bibinfo {author} {\bibfnamefont {R.~A.}\ \bibnamefont {Soref}},
  \ and\ \bibinfo {author} {\bibfnamefont {J.~A.}\ \bibnamefont {{Del
  Alamo}}},\ }\href {\doibase 10.1109/3.44924} {\bibfield  {journal} {\bibinfo
  {journal} {IEEE J. Quant. Electron.}\ }\textbf {\bibinfo {volume} {26}},\
  \bibinfo {pages} {113} (\bibinfo {year} {1990})}\BibitemShut {NoStop}%
\bibitem [{\citenamefont {Ghimire}\ \emph {et~al.}(2012)
	\citenamefont {Ghimire},
	\citenamefont {Dichiara},
	\citenamefont {Sistrunk},
	\citenamefont {Ndabashimiye},
	\citenamefont {Szafruga},
	\citenamefont {Mohammad},
	\citenamefont {Agostini},
	\citenamefont {Dimauro},
	\ and\ \citenamefont {Reis}}]{Ghimire2012}%
\BibitemOpen
\bibfield  {author} {
	\bibinfo {author} {\bibfnamefont {S.}\ \bibnamefont {Ghimire}},
	\bibinfo {author} {\bibfnamefont {A.~D.}\ \bibnamefont {Dichiara}},
	\bibinfo {author} {\bibfnamefont {E.}\ \bibnamefont {Sistrunk}},
	\bibinfo {author} {\bibfnamefont {G.}\ \bibnamefont {Ndabashimiye}},
	\bibinfo {author} {\bibfnamefont {U.~B.}\ \bibnamefont {Szafruga}},
	\bibinfo {author} {\bibfnamefont {A.}\ \bibnamefont {Mohammad}},
	\bibinfo {author} {\bibfnamefont {P.}\ \bibnamefont {Agostini}},
	\bibinfo {author} {\bibfnamefont {L.~F.}\ \bibnamefont {Dimauro}},
	\ and\ \bibinfo {author} {\bibfnamefont {D.~A.}\ \bibnamefont {{Reis}}},\ }\href {\doibase 10.1103/PhysRevA.85.043836} {\bibfield  {journal} {\bibinfo
		{journal} {Phys. Rev. A}\ }\textbf {\bibinfo {volume} {85}},\
	\bibinfo {pages} {043836} (\bibinfo {year} {2012})}\BibitemShut {NoStop}%
\bibitem [{\citenamefont {Liu}\ \emph {et~al.}(2018{\natexlab{b}})\citenamefont
  {Liu}, \citenamefont {Powell}, \citenamefont {Zarate},\ and\ \citenamefont
  {Shadrivov}}]{Liu2018g}%
  \BibitemOpen
  \bibfield  {author} {\bibinfo {author} {\bibfnamefont {M.}~\bibnamefont
  {Liu}}, \bibinfo {author} {\bibfnamefont {D.~A.}\ \bibnamefont {Powell}},
  \bibinfo {author} {\bibfnamefont {Y.}~\bibnamefont {Zarate}}, \ and\ \bibinfo
  {author} {\bibfnamefont {I.~V.}\ \bibnamefont {Shadrivov}},\ }\href {\doibase
  10.1103/PhysRevX.8.031077} {\bibfield  {journal} {\bibinfo  {journal} {Phys.
  Rev. X}\ }\textbf {\bibinfo {volume} {8}},\ \bibinfo {pages} {31077}
  (\bibinfo {year} {2018}{\natexlab{b}})}\BibitemShut {NoStop}%
\end{thebibliography}
\end{document}